\documentclass[%
 reprint, 
 amsmath,amssymb,
 aps, physrev,
]{revtex4-2}

\usepackage{graphicx}
\usepackage{dcolumn}
\usepackage{bm}
\usepackage{hyperref}
\hypersetup{colorlinks=true, urlcolor=blue, linkcolor=blue, citecolor=blue}

\begin{abstract}

Flying insect swarms exhibit cohesion without the local velocity alignment observed in flocks of birds or schools of fish. The interaction rules and channels of communication between insects that lead to collective behavior have not yet been determined. We propose a theoretical model based on acoustic communication between swarm members, where each individual is attracted to a weighted mean field of its neighbors. We demonstrate that this simple framework can describe a wide range of swarming dynamics, including a phenomenon where pairs of insects break free from the swarm in a helical orbit around each other. Counterintuitively, our numerical model suggests that stochastic noise enhances the stability of an insect swarm, as it interferes with pair formation. Finally, we demonstrate that a specific species of malarial mosquito produces swarms that reside on the edge of a transition to instability. Our study shows that fairly simple local interaction rules can be sufficient to describe the collective behavior of swarming biological systems.

\end{abstract}

\begin{document}

\title{\textbf{Noise-induced stability of insect swarms}}

\author{Justin Faber}
\email{faber@physics.ucla.edu}
\affiliation{Department of Physics \& Astronomy, University of California, Los Angeles, California, USA}

\author{Annabelle Boots}
\affiliation{Department of Physics \& Astronomy, University of California, Los Angeles, California, USA}

\author{Dolores Bozovic}
\affiliation{Department of Physics \& Astronomy, University of California, Los Angeles, California, USA}
\affiliation{California NanoSystems Institute, University of California, Los Angeles, California, USA}

\date{\today}

\maketitle

Bird, fish, and insect swarms have long been sources of fascination, with open questions about how they arise, morph over time, and lead to collective motion \cite{garnierBiologicalPrinciplesSwarm2007, sumpterPrinciplesCollectiveAnimal2005, bernoffNonlocalAggregationModels2013, ouellettePhysicsPerspectiveCollective2022}. These ``superorganisms'' form not as a result of external forces or instruction by a leader, but through local interaction rules between individual group members. The vast variety of collective phenomena that have been observed in nature, therefore, must arise from variations in these local interactions. Early numerical models \cite{reynoldsFlocksHerdsSchools1987} and many others that followed have considered some combination of three basic interaction rules \cite{vicsekNovelTypePhase1995, ginelliPhysicsVicsekModel2016, carruiteroInertialSpinModel2024}. First, each agent experiences an attractive force to its neighbors, with varying dependence on metric or topological distances. Second, pairs of agents experience repulsive forces at short distances, thus avoiding collisions with each other and the collapse of an entire aggregation. Third, each agent attempts to align its velocity with that of its neighbors. These simple interaction rules, combined with active motion of individual agents, as well as random fluctuations due to the external environment, produce many different states of emergent order \cite{gonzalez-albaladejoMeanfieldTheoryChaotic2023, dorsognaSelfPropelledParticlesSoftCore2006} and were shown to describe the behavior of birds \cite{balleriniInteractionRulingAnimal2008}, fish \cite{huthSimulationMovementFish1992}, insects \cite{gorbonosLongrangeAcousticInteractions2016, gonzalez-albaladejoMeanfieldTheoryChaotic2023}, bacteria \cite{vicsekCollectiveMotion2012}, and many other biological superorganisms.

Mosquitoes and the dynamics of their swarming behavior are of particular interest due to their detrimental public health impact. These insects and the pathogens they spread are responsible for over 700,000 human deaths each year and impose immense financial burden on the healthcare systems of mosquito-endemic regions \cite{WHO2018, WHO2020}. Further, mosquitoes grow resistant to the pesticides we develop, and the habitats of dangerous species are expanding due to increasing global temperatures \cite{messinaManyProjectedFutures2015}. The need for additional population control strategies is therefore clear and ever-increasing.

One promising strategy for population control is to interfere with the mosquito's mating behavior \cite{suAssessingAcousticBehaviour2020, andresBuzzkillTargetingMosquito2020}. At dusk, male mosquitoes form swarms, which attract females. When a female enters the swarm, males detect the acoustic buzzing of her beating wings and pursue her for a chance to mate. If swarm formation could be interfered with, acoustically or otherwise, mating behavior among mosquitoes would be disrupted, yielding a means of population control for dangerous species. However, our understanding of mosquito swarming dynamics is currently limited. We do not know the interaction rules between mosquitoes or by what means they exchange information (visual, auditory, olfactory, or a combination thereof). Further, we do not know what controls the behavioral change that results in swarm formation.

In this letter, we propose a numerical model to describe the dynamics of mosquito swarming. As a simplification, we consider only the acoustic form of interaction between individual insects. We note however that the near-field acoustics of flying insects is in itself non-trivial \cite{seoMechanismScalingWing2019, faberLocalizationFrequencyResolution2026, sueurSoundRadiationFlying2005}. How rapidly signal strength falls off with distance depends on how well the near-field approximation holds, whether the insect detects the pressure field or the velocity field of sound, and which terms in the multipole expansion are strongest for the sound sources. To capture this complexity in a single parameter, we allow the acoustic signal strength to fall off as $1/r^{\gamma}$, where $r$ is the distance between two insects and $\gamma$ is a free parameter in our model. Based on previous studies of the mosquito auditory system, we expect $\gamma$ to be between $1$ and $3$ \cite{faberLocalizationFrequencyResolution2026}. The numerical model we propose is entirely based on local attraction, where each insect accelerates in the direction of the weighted mean field of its neighbors, with $\gamma$ controlling the weighting of this mean field. We make no assumption of velocity alignment or repulsive forces, as these are not warranted by experimental observations \cite{cribellierComplexSwarmingDynamics2024} (see Supplementary Material). Inertia of the insects' flight paths was found to be sufficient to prevent the swarm collapse.

We introduce stochastic noise into the flight path of each individual insect and explore the stability of the swarm throughout the $(\gamma, \eta)$ parameter space, where $\eta$ is the noise strength. Surprisingly, we find that intermediate levels of noise enhance swarm stability, over a rough range of $4>\gamma>2$. Our model further predicts the occurrence of pair formation, an event whereby pairs of insects break free from the swarm in a helical orbit around each other. This phenomenon has been observed in experimental data of midge swarms \cite{puckettTimeFrequencyAnalysisReveals2015}. Finally, we show how experimental swarming data can be mapped onto this model. We demonstrate this approach on swarming data of the malaria mosquito from prior literature \cite{podaSpatialTemporalCharacteristics2024} and identify the point in $(\gamma, \eta)$ parameter space where each swarm resides. Interestingly, the experimental swarms map onto stable points near the edge of the phase transition between stable and unstable swarming behavior. We speculate that this regime is biologically favorable, as it allows swarm members to easily break free from the swarm in order to pursue potential mates or evade predators.

\begin{figure}[t!]
\includegraphics[width=\columnwidth]
{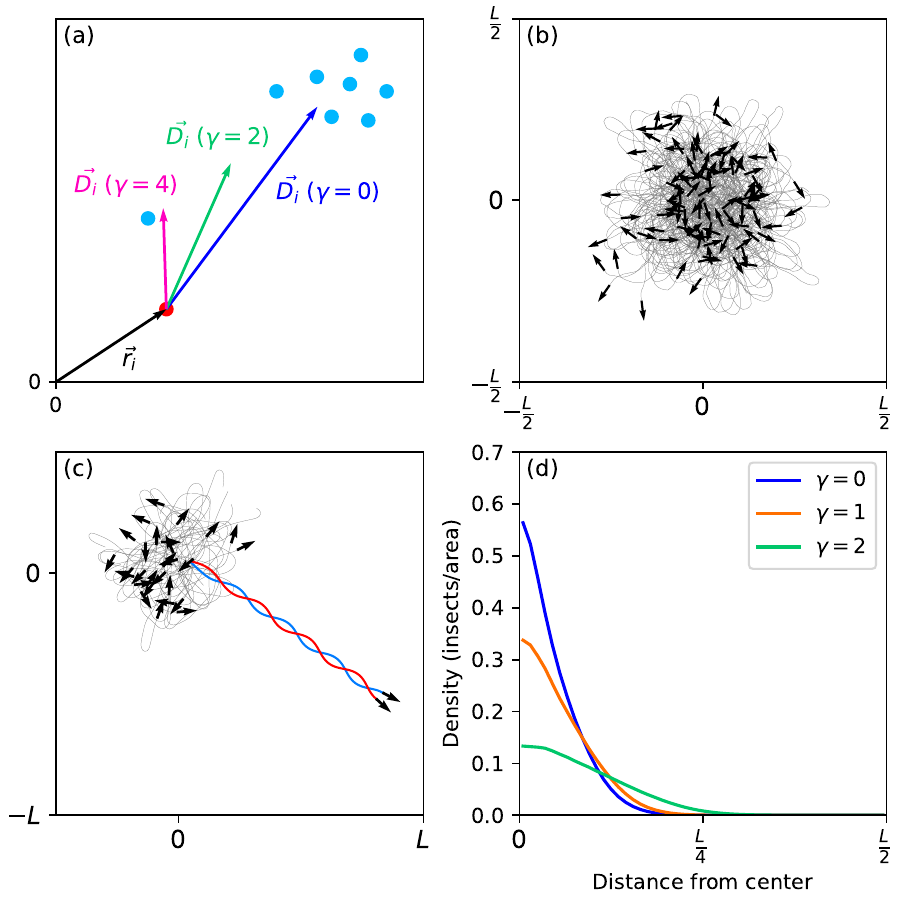}
\caption{(a) Illustration of how the weighted mean-field vector, $\vec{D}_i$, depends on $\gamma$. The reference insect is depicted as a red dot, while blue dots represent neighboring insects. (b) A snapshot of a stable swarm with $N=100$, $\gamma = 2$, and $\eta = 0.1$. Black vectors indicate the positions and angles of orientations of the insects. Path flights are traced with grey curves. (c) An instance of a pair formation event, shown by red and blue traces, from a swarm with $N=30$, $\gamma = 3$, and $\eta = 0$. (d) Time-average density profiles of swarms with various $\gamma$ values ($N=100$ and $\eta = 0.1$).}
\label{fig-Intro}
\end{figure}

\textit{Numerical model}---The attractive force between an individual insect and other members of the swarm follows Newton's second law,

\begin{eqnarray}
m\frac{d\vec{v}_i}{dt} = k \vec{D}_i = k \frac{ \sum_{j\neq i} \vec{r}_{ij}/r_{ij}^\gamma }{\sum_{j\neq i} 1/r_{ij}^\gamma },
\label{eq:mv_dot}
\end{eqnarray}

\noindent where $\vec{v}_i$ is the velocity vector of the $i^{th}$ insect, $m$ is its effective mass, and $k$ is an effective spring constant characterizing the strength of the attraction to the weighted mean field. The vector $\vec{D}_i$ points from the $i^{th}$ insect to the weighted mean field of its neighbors, where $\vec{r}_{ij}$ is the vector pointing from the $i^{th}$ to the $j^{th}$ insect. The denominator of $\vec{D}_i$ scales the weighted mean field to make the attractive force Hookean. A scaling similar to this was originally introduced in an adaptive-gravity model of insect swarms \cite{gorbonosLongrangeAcousticInteractions2016}. The mean-field weighting exponent, $\gamma$, characterizes how much priority is given to the nearest few neighbors. For $\gamma=0$, all neighbors are equally weighted, while for higher $\gamma$ values, nearer neighbors provide more influence than farther neighbors (Fig. \ref{fig-Intro} (a)). In the limit of $\gamma \to \infty$, each insect will be attracted only to its nearest neighbor.

We introduce the exponent, $\gamma$, to capture how rapidly positional information falls off with distance. If the sound generated by an insect displays the characteristics of an acoustic dipole, and their auditory organs detect the particle velocity field produced by the wingbeats, then we expect the interaction term to fall of as $1/r^3$ in the near-field approximation and should use $\gamma=3$ \cite{faberLocalizationFrequencyResolution2026, sueurSoundRadiationFlying2005}. However, for insects that detect each others' velocity fields from monopole sources or pressure fields from dipole sources, we expect the signal to fall off as $1/r^2$ \cite{bennet-clarkSizeScaleEffects1998}. Lastly, if we are describing large animals that detect the pressure field of sound and whose acoustic fields can be approximated as monopole sources, then the acoustic signal should fall of as $1/r$. As insects exhibit diverse means of acoustic communication \cite{gopfertHearingInsects2016, romerDirectionalHearingInsects2020}, this parameter allows the model to be adapted for a wide range of applications. Further, the means of communication can be inferred from matching the statistics of the model to those of experimental data (see Supplementary Material).

We now introduce the constraint of constant speed, consistent with assumptions used in the classic Vicsek model \cite{vicsekNovelTypePhase1995}. We can represent the velocity vector as $\vec{v}_i = v_0 \hat{v}_i$, where $v_0 = |\vec{v}_i|$ is the constant speed for all insects at all times. To simplify the equations of motion, we first split the weighted mean-field vector into components that are parallel and perpendicular to $\vec{v}_i$: $\vec{D}_i = \vec{D}_{i \parallel} + \vec{D}_{i \bot}$. By the assumption of constant speed, we can neglect the parallel component and write the equations of motion as

\begin{eqnarray}
\frac{m v_0}{k}\frac{d\hat{v}_i}{dt} =  \vec{D}_{i \bot} = \vec{D}_i - (\vec{D}_i \cdot \hat{v}_i)\hat{v}_i,
\end{eqnarray}

\noindent where the right side of the equation is the vector rejection of $\vec{D}_{i}$ onto $\hat{v}_i$.

Previous experimental measurements of malaria mosquito swarms have shown that the dynamics occur predominantly in planes perpendicular to the direction of gravity, with little velocity and acceleration measured in the $\hat{z}$ direction  \cite{cavagnaCharacterizationLabbasedSwarms2023, podaSpatialTemporalCharacteristics2024}. Therefore, we neglect this component in our numerical model and describe the dynamics in just two dimensions. We introduce this constraint by writing the weighted mean-field vector as $\vec{D}_i = D_{i,x} \hat{x} + D_{i,y} \hat{y}$, and expressing the the velocity vector as $\hat{v}_i = \cos\theta_i \hat{x} + \sin\theta_i \hat{y}$. Making these substitutions and expressing the differential equations as difference equations, we find

\begin{eqnarray}
\theta_i(t + \Delta t) = \theta_i(t) + \frac{k \Delta t}{m v_0}\bigg[D_{i,y}(t)\cos\theta_i(t) - D_{i,x}(t)\sin\theta_i(t) \bigg] \notag
\end{eqnarray}

\begin{eqnarray}
\vec{r}_i(t + \Delta t) = \vec{r}_i(t) + v_0 \Delta t \hat{v}_i(t),
\end{eqnarray}

\noindent where $\vec{r}_i$ and $\theta_i$ represent the position and orientation angle, respectively, of the $i^{th}$ agent. The system can be characterized by two length scales. The short length scale is given by $\ell = v_0 \Delta t$, which quantifies how much each agent moves with one time step, $\Delta t$. We represent the large length scale as $L = \pi\frac{m v_0}{k \Delta t}$, which determines the upper bound on the dynamics. When an agent is at a distance $L$ from the weighted mean field, it is able to rotate an angle of $\pi$ in a single time step. At separations greater than L, we expect the system to exhibit edge effects. To minimize these effects, we rectify the maximum rotation angle to be $\pm\pi$ in any given time step.

We introduce noise into our system in a manner consistent with prior studies. At each time step, a random angle selected from a uniform probability distribution is added to the orientation of each agent. We now present the model in its final form, which can readily be simulated:

\begin{eqnarray}
\begin{split}
\theta_i(t + \Delta t) = \theta_i(t) + \frac{\pi}{L}\Big[D_{i,y}(t)\cos\theta_i(t) &- D_{i,x}(t)\sin\theta_i(t) \Big] \\ &+ \eta \pi U[-1, 1] \notag
\end{split}
\end{eqnarray}

\vspace{-1 mm}

\begin{eqnarray}
\vec{r}_i(t + \Delta t) = \vec{r}_i(t) + \ell \hat{v}_i(t)
\label{eq:theta_diff}
\end{eqnarray}

\vspace{-3 mm}

\begin{eqnarray}
\vec{D}_i =  \frac{ \sum_{j\neq i} \vec{r}_{ij}/r_{ij}^\gamma }{\sum_{j\neq i} 1/r_{ij}^\gamma }, \notag
\end{eqnarray}

\noindent where $U[-1, 1]$ is an instance of a uniform random variable in the range $[-1, 1]$, and $\eta$ characterizes the noise strength, ranging from $[0, 1]$. Without loss of generality, we set $\ell=1$. The upper length scale, $L$, determines both the length scale of the swarm and the time step. The model requires the constraint that $L \gg \ell$. Unless otherwise stated, we use $L=100$. For greater values of $L$, we observe a trivial rescaling of the swarm size and excessive sampling of the insect trajectories (see Supplementary Material). Throughout this letter, we vary the remaining three parameters: the number of insects ($N$), the mean-field weighting exponent ($\gamma$), and the noise strength ($\eta$).

\textit{Pair formation}---For a wide range of parameters, we observe that stable swarms emerge, regardless of the initial conditions of the agents (Fig. \ref{fig-Intro} (b)). For sufficiently high values of $\gamma$, swarms develop but then slowly disintegrate through the occurrence of pair formation, whereby pairs of insects exit the swarm while orbiting each other in a spiral pattern (Fig. \ref{fig-Intro} (c)). This behavior has been observed in midge swarms \cite{puckettTimeFrequencyAnalysisReveals2015} and was proposed to arise through an adaptive-gravity mechanism \cite{gorbonosPairFormationInsect2020}. Our model also captures this phenomenon in the $\gamma>2$ regime, in which the attraction to each agent's nearest neighbor dominates the direction of the weighted mean-field vector, $\vec{D}_i$. Consistent with experimental data \cite{kelleyEmergentDynamicsLaboratory2013} (see Supplementary Material), we find that the insect density profile follows a Gaussian-like distribution, which broadens for increasing $\gamma$ (Fig. \ref{fig-Intro} (d)).

\begin{figure}[t!]
\includegraphics[width=\columnwidth]{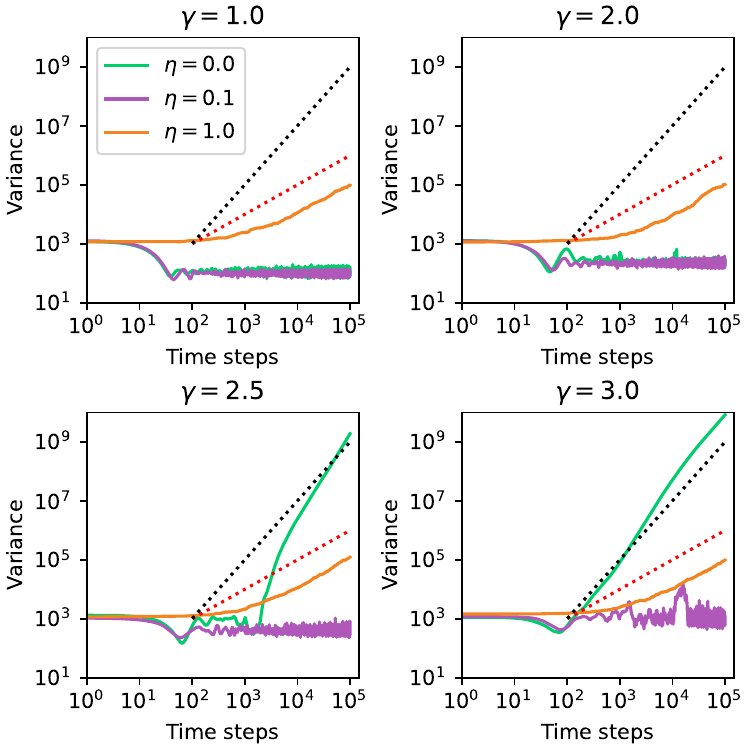}
\caption{Positional variance growth as a function of time for several noise strengths ($\eta$) and mean-field weighting exponents ($\gamma$). Stochastic noise has a stabilizing effect on the swarms, roughly in the $4>\gamma>2$ regime. Linear and quadratic growth are shown with dotted red and black lines, respectively. For all simulations, $N=100$.}
\label{fig-variance}
\end{figure}

\textit{Swarm stability analysis}---To characterize the stability of a swarm, we plot the radial variance as a function of time and fit a power law: 

\begin{eqnarray}
\frac{1}{N} \sum_{i=1}^N |\vec{r}_i(t) - \vec{r}_{cm}(t)|^2 \sim t^\alpha, 
\end{eqnarray}

\noindent where $\vec{r}_{cm}(t) = \frac{1}{N} \sum_{i=1}^N \vec{r}_i(t)$ is the swarm center of mass. $\alpha \approx 0$ identifies a stable swarm, while $\alpha \approx 1$ indicates diffusive behavior, and $\alpha \approx 2$ indicates ballistic motion. We plot the radial variance for several choices of ($\gamma$, $\eta$) parameters and observe all three regimes (Fig. \ref{fig-variance}). Counterintuitively, we observe that, for roughly $4>\gamma>2$, the presence of stochastic noise is essential for the formation of stable swarms.

The stabilizing effect of noise is reminiscent of the effect of stochastic resonance \cite{wellensStochasticResonance2003}, and has previously been proposed to give rise to coherent motion within flocks of birds \cite{reynoldsStochasticityMayGenerate2023}. In our numerical model, this phenomenon can be understood as follows. For sufficiently high $\gamma$, The attractive force on each agent will be dominated by few of its nearest neighbors, resulting in the formation of pairs or small groups, which can break free from the swarm. However, sufficiently high levels of noise can cause the pair members to diverge, thus breaking the pair bond and allowing the two insects to be absorbed back into the swarm. In the Appendix, we provide an analytic calculation that approximates the pair formation and annihilation rate. Equating the two allows us to estimate the boundary between stable and unstable swarming dynamics.

At very high noise strengths ($\eta \approx 1$), a swarm can lose stability, not through pair formation, but through diffusion of the entire swarm. In the Supplementary Material, we also estimate the boundary for this transition. In both the analytic calculation and the numerical simulations, the boundary is remarkably close to $\eta =1$, indicating that stable swarms can still exist even when the agents' dynamics are dominated by noise. In this regime, the agents' dynamics are random walks with weak biases toward their neighbors.

\begin{figure}[t!]
\includegraphics[width=\columnwidth]{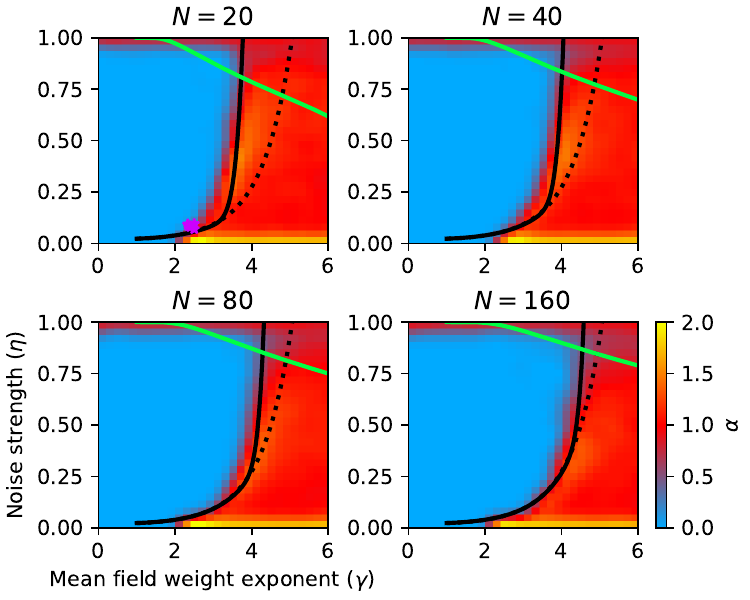}
\caption{Phase diagrams in ($\gamma$, $\eta$) space for several swarm sizes ($N$). Stable swarms reside in the blue, $\alpha \approx 0$ region. Solid black and green curves show the analytic calculations of the lower and upper phase transitions (see Appendix and Supplementary Material). The dotted black curves show the analytic calculation of the lower phase transition in the limit of $N \to \infty$. Purple stars identify parameter values of six experimental mosquito swarm datasets when mapped onto the model.}
\label{fig-phase_diag}
\end{figure}

To further explore the dynamics of swarm formation, we compute the full phase diagram in ($\gamma$, $\eta$) space for several swarm sizes (Fig. \ref{fig-phase_diag}). For $\gamma<2$, swarms are stable for nearly any level of noise. In fact, swarm stability persists even when the dynamics are almost entirely determined by the stochastic term. For intermediate weighting exponents (roughly $4>\gamma>2$), noise is essential for swarm stability. The analytic calculations confirm our hypothesis that noise can stabilize a swarm by destabilizing insect pair bonds. Finally, for roughly $\gamma>4$, all swarms are unstable. Both the numerical simulations and the analytic calculations indicate that the region of stability in ($\gamma$, $\eta$) space slightly expands with increasing swarm size ($N$). For comparison, we also show the analytic calculation in the $N \to \infty$ limit.

\textit{Comparison to experimental data}---Apart from providing a general framework for swarm formation, our numerical model can be used to infer the spatial dependence of the interaction between agents. By matching swarm statistics of the model to those of the experimental data, we can estimate where in ($\gamma$, $\eta$) space the data reside. As a demonstration, we use previously published data obtained from swarms of malaria mosquitoes \cite{podaSpatialTemporalCharacteristics2024}. We determine the correlation length of individual mosquito trajectories and the angular diffusion constant over short length scales. For an interpretable comparison to the numerical model, we nondimensionalize these measures, using the mean swarm radius and the mean speed of individual agents (see Supplementary Material). 

Comparing the experimental data to the numerical model, we find $\gamma = 2.4 \pm 0.1$ and $\eta = 0.08 \pm 0.01$ (mean and standard deviation across six swarms). Such a value for $\gamma$ could arise from a combination of strong monopole and dipole terms in the radiation of sound by flying mosquitoes. Alternatively, the near-field approximation may not be satisfied, resulting in a velocity field that falls off faster than $1/r^2$ but slower than $1/r^3$. We note that $\gamma \approx 2.4$ is in good agreement with experimental measurements of the velocity fields produced by the wings of tethered mosquitoes \cite{arthurMosquitoAedesAegypti2014}.

\textit{Discussion}---We have developed a theoretical model capable of describing the formation and stability of insect swarms. The general framework is also applicable to other examples of collective phenomena in biology, in which local attraction provides the dominant interaction. We followed upon the framework established by the Vicsek model, but with attraction as the only interaction rule, rather than velocity alignment. With the same number of free parameters, this novel system captures the key features of swarming behavior exhibited by flying insects. 

Insect pair formation has been observed experimentally in swarms of midges \cite{puckettTimeFrequencyAnalysisReveals2015}, where two insects form a tight helical orbit around each other, displaying nearly sinusoidal trajectories. This phenomenon has been captured by an adaptive gravity model when adaptive sensing is included \cite{gorbonosPairFormationInsect2020}. In the numerical model we propose, pair formation arises naturally in the $\gamma > 2$ regime due to high attraction to nearest neighbors. We note that in the absence of noise, pairs can remain stable indefinitely, while in the presence of noise, pair formation occurs in a transient manner, as observed experimentally  \cite{puckettTimeFrequencyAnalysisReveals2015}.

With sufficient occurrence of pair formation, the swarm becomes unstable and disintegrates. The phase diagrams presented in this letter suggest that swarming cannot occur if the interaction between individual members falls off too rapidly with distance (approximately $\gamma > 4$). The maximum value for $\gamma$ depends on the number of agents present, with larger swarms exhibiting a larger stable regime. We also found that the presence of noise is essential for stable swarming in the intermediate regime (approximately $4 >\gamma > 2$). 

As acoustic interference provides a potential strategy for population control in mosquitoes \cite{andresBuzzkillTargetingMosquito2020}, it is beneficial to explore how the presence of noise impacts swarm formation and stability. Our results indicate that introducing stochastic noise serves to enhance rather than weaken the stability of the swarm. As swarm stability persists even when the dynamics are dominated by noise ($\eta > 0.9$), we predict that using broadband noise to disrupt insect swarms would be an ineffective strategy. Instead, we speculate that effective strategies will include sound composed of just one, or a few tones at any given time, possibly with dynamic frequency sweeps \cite{faberSoundMaskingStrategies2026}. Future work entails using this model to test and compare candidate signals for swarm disruption.

The mean-field weighting exponent ($\gamma$) can be varied, allowing this model to be adapted to describe a diverse range of biological systems. Agents that communicate acoustically through the pressure or velocity fields of sound, and whose acoustic field is dominated by any term in the multipole expansion \cite{faberLocalizationFrequencyResolution2026}, can readily be captured through this parameter. We note that higher integer values of this parameter have been studied in the context of swarm collapse \cite{gorbonosStableSwarmingUsing2017}. Further, information about the channels of communication between agents can be inferred by fitting this parameter to experimental data, as demonstrated in this letter.

Both visual and auditory cues have been shown to be relevant in collision avoidance between mosquitoes \cite{guptaMosquitoesIntegrateVisual2024}. Assuming sufficient lighting, visual information should not fall off with distance in these sparse swarms. Further, long-range pheromonal attraction of female to male mosquitoes appears to be absent \cite{podaNoEvidenceLongrange2022}. We therefore reason that an auditory interaction is the most plausible channel of communication between swarming mosquitoes. This hypothesis is supported by the importance of the mosquito auditory system for locating potential mates \cite{gopfertHearingInsects2016, faberMosquitoinspiredTheoreticalFramework2025a}, the observed male-male acoustic interactions in tethered mosquitoes \cite{aldersleyEmergentAcousticOrder2017}, and the male-male flight interactions within swarms \cite{iacomelliExperimentalEvidenceMale2026}. Further, the mean-field weighting exponent of $\gamma \approx 2.4$ can be fully explained by an acoustic interaction \cite{faberLocalizationFrequencyResolution2026}.

Interestingly, we found that mosquito swarms reside on the verge of instability. This finding is consistent with conclusions from previous studies, which show that swarming midges are only weakly bound to the swarm \cite{puckettSearchingEffectiveForces2014}. Weak stability would allow for rapid transitions between swarming and non-swarming states, with minimal adjustments of individual behavior. We speculate that this would be advantageous to insects swarms, as it would allow individuals to quickly break free from the swarm to pursue a potential mate or escape the attack of a predator. Furthermore, the edge of instability may offer the advantage of long-range information transfer through the swarm. Computation at the edge of criticality has been a topic of extensive interest and debate in neuroscience \cite{obyrneHowCriticalBrain2022}, condensed matter physics \cite{paschenQuantumPhasesDriven2021}, machine learning \cite{bahriStatisticalMechanicsDeep2020}, and complex systems \cite{watkins25YearsSelforganized2016}. Future work entails exploring its role in collective behavior exhibited by a swarm.

\vspace{1mm}

\textit{Acknowledgments}---This work was funded in part by NSF Physics of Living Systems.

\vspace{1mm}

\textit{Data availability}---All computer code used for analysis is publicly available online \cite{DataAvailability}.

\bibliography{Bibliography}

\begin{thebibliography}{47}%
\makeatletter
\providecommand \@ifxundefined [1]{%
 \@ifx{#1\undefined}
}%
\providecommand \@ifnum [1]{%
 \ifnum #1\expandafter \@firstoftwo
 \else \expandafter \@secondoftwo
 \fi
}%
\providecommand \@ifx [1]{%
 \ifx #1\expandafter \@firstoftwo
 \else \expandafter \@secondoftwo
 \fi
}%
\providecommand \natexlab [1]{#1}%
\providecommand \enquote  [1]{``#1''}%
\providecommand \bibnamefont  [1]{#1}%
\providecommand \bibfnamefont [1]{#1}%
\providecommand \citenamefont [1]{#1}%
\providecommand \href@noop [0]{\@secondoftwo}%
\providecommand \href [0]{\begingroup \@sanitize@url \@href}%
\providecommand \@href[1]{\@@startlink{#1}\@@href}%
\providecommand \@@href[1]{\endgroup#1\@@endlink}%
\providecommand \@sanitize@url [0]{\catcode `\\12\catcode `\$12\catcode
  `\&12\catcode `\#12\catcode `\^12\catcode `\_12\catcode `\%12\relax}%
\providecommand \@@startlink[1]{}%
\providecommand \@@endlink[0]{}%
\providecommand \url  [0]{\begingroup\@sanitize@url \@url }%
\providecommand \@url [1]{\endgroup\@href {#1}{\urlprefix }}%
\providecommand \urlprefix  [0]{URL }%
\providecommand \Eprint [0]{\href }%
\providecommand \doibase [0]{https://doi.org/}%
\providecommand \selectlanguage [0]{\@gobble}%
\providecommand \bibinfo  [0]{\@secondoftwo}%
\providecommand \bibfield  [0]{\@secondoftwo}%
\providecommand \translation [1]{[#1]}%
\providecommand \BibitemOpen [0]{}%
\providecommand \bibitemStop [0]{}%
\providecommand \bibitemNoStop [0]{.\EOS\space}%
\providecommand \EOS [0]{\spacefactor3000\relax}%
\providecommand \BibitemShut  [1]{\csname bibitem#1\endcsname}%
\let\auto@bib@innerbib\@empty
\bibitem [{\citenamefont {Garnier}\ \emph {et~al.}(2007)\citenamefont
  {Garnier}, \citenamefont {Gautrais},\ and\ \citenamefont
  {Theraulaz}}]{garnierBiologicalPrinciplesSwarm2007}%
  \BibitemOpen
  \bibfield  {author} {\bibinfo {author} {\bibfnamefont {S.}~\bibnamefont
  {Garnier}}, \bibinfo {author} {\bibfnamefont {J.}~\bibnamefont {Gautrais}},\
  and\ \bibinfo {author} {\bibfnamefont {G.}~\bibnamefont {Theraulaz}},\
  }\bibfield  {title} {\bibinfo {title} {The biological principles of swarm
  intelligence},\ }\href {https://doi.org/10.1007/s11721-007-0004-y} {\bibfield
   {journal} {\bibinfo  {journal} {Swarm Intell.}\ }\textbf {\bibinfo {volume}
  {1}},\ \bibinfo {pages} {3} (\bibinfo {year} {2007})}\BibitemShut {NoStop}%
\bibitem [{\citenamefont
  {Sumpter}(2005)}]{sumpterPrinciplesCollectiveAnimal2005}%
  \BibitemOpen
  \bibfield  {author} {\bibinfo {author} {\bibfnamefont {D.}~\bibnamefont
  {Sumpter}},\ }\bibfield  {title} {\bibinfo {title} {The principles of
  collective animal behaviour},\ }\href
  {https://doi.org/10.1098/rstb.2005.1733} {\bibfield  {journal} {\bibinfo
  {journal} {Philos. Trans. R. Soc. B}\ }\textbf {\bibinfo {volume} {361}},\
  \bibinfo {pages} {5} (\bibinfo {year} {2005})}\BibitemShut {NoStop}%
\bibitem [{\citenamefont {Bernoff}\ and\ \citenamefont
  {Topaz}(2013)}]{bernoffNonlocalAggregationModels2013}%
  \BibitemOpen
  \bibfield  {author} {\bibinfo {author} {\bibfnamefont {A.~J.}\ \bibnamefont
  {Bernoff}}\ and\ \bibinfo {author} {\bibfnamefont {C.~M.}\ \bibnamefont
  {Topaz}},\ }\bibfield  {title} {\bibinfo {title} {Nonlocal {{Aggregation
  Models}}: {{A Primer}} of {{Swarm Equilibria}}},\ }\href
  {https://doi.org/10.1137/130925669} {\bibfield  {journal} {\bibinfo
  {journal} {SIAM Rev.}\ }\textbf {\bibinfo {volume} {55}},\ \bibinfo {pages}
  {709} (\bibinfo {year} {2013})}\BibitemShut {NoStop}%
\bibitem [{\citenamefont
  {Ouellette}(2022)}]{ouellettePhysicsPerspectiveCollective2022}%
  \BibitemOpen
  \bibfield  {author} {\bibinfo {author} {\bibfnamefont {N.~T.}\ \bibnamefont
  {Ouellette}},\ }\bibfield  {title} {\bibinfo {title} {A physics perspective
  on collective animal behavior},\ }\href
  {https://doi.org/10.1088/1478-3975/ac4bef} {\bibfield  {journal} {\bibinfo
  {journal} {Phys. Biol.}\ }\textbf {\bibinfo {volume} {19}},\ \bibinfo {pages}
  {021004} (\bibinfo {year} {2022})}\BibitemShut {NoStop}%
\bibitem [{\citenamefont {Reynolds}(1987)}]{reynoldsFlocksHerdsSchools1987}%
  \BibitemOpen
  \bibfield  {author} {\bibinfo {author} {\bibfnamefont {C.~W.}\ \bibnamefont
  {Reynolds}},\ }\bibfield  {title} {\bibinfo {title} {Flocks, herds and
  schools: {{A}} distributed behavioral model},\ }in\ \href
  {https://doi.org/10.1145/37401.37406} {\emph {\bibinfo {booktitle}
  {Proceedings of the 14th Annual Conference on {{Computer}} Graphics and
  Interactive Techniques}}},\ \bibinfo {series and number} {{{SIGGRAPH}} '87}\
  (\bibinfo  {publisher} {Association for Computing Machinery},\ \bibinfo
  {year} {1987})\ pp.\ \bibinfo {pages} {25--34}\BibitemShut {NoStop}%
\bibitem [{\citenamefont {Vicsek}\ \emph {et~al.}(1995)\citenamefont {Vicsek},
  \citenamefont {Czir{\'o}k}, \citenamefont {{Ben-Jacob}}, \citenamefont
  {Cohen},\ and\ \citenamefont {Shochet}}]{vicsekNovelTypePhase1995}%
  \BibitemOpen
  \bibfield  {author} {\bibinfo {author} {\bibfnamefont {T.}~\bibnamefont
  {Vicsek}}, \bibinfo {author} {\bibfnamefont {A.}~\bibnamefont {Czir{\'o}k}},
  \bibinfo {author} {\bibfnamefont {E.}~\bibnamefont {{Ben-Jacob}}}, \bibinfo
  {author} {\bibfnamefont {I.}~\bibnamefont {Cohen}},\ and\ \bibinfo {author}
  {\bibfnamefont {O.}~\bibnamefont {Shochet}},\ }\bibfield  {title} {\bibinfo
  {title} {Novel {{Type}} of {{Phase Transition}} in a {{System}} of
  {{Self-Driven Particles}}},\ }\href
  {https://doi.org/10.1103/PhysRevLett.75.1226} {\bibfield  {journal} {\bibinfo
   {journal} {Phys. Rev. Lett.}\ }\textbf {\bibinfo {volume} {75}},\ \bibinfo
  {pages} {1226} (\bibinfo {year} {1995})}\BibitemShut {NoStop}%
\bibitem [{\citenamefont {Ginelli}(2016)}]{ginelliPhysicsVicsekModel2016}%
  \BibitemOpen
  \bibfield  {author} {\bibinfo {author} {\bibfnamefont {F.}~\bibnamefont
  {Ginelli}},\ }\bibfield  {title} {\bibinfo {title} {The {{Physics}} of the
  {{Vicsek}} model},\ }\href {https://doi.org/10.1140/epjst/e2016-60066-8}
  {\bibfield  {journal} {\bibinfo  {journal} {Eur. Phys. J. Spec. Top.}\
  }\textbf {\bibinfo {volume} {225}},\ \bibinfo {pages} {2099} (\bibinfo {year}
  {2016})}\BibitemShut {NoStop}%
\bibitem [{\citenamefont {Carruitero}\ \emph {et~al.}(2024)\citenamefont
  {Carruitero}, \citenamefont {Duran}, \citenamefont {Pisegna}, \citenamefont
  {Sturla},\ and\ \citenamefont {Grigera}}]{carruiteroInertialSpinModel2024}%
  \BibitemOpen
  \bibfield  {author} {\bibinfo {author} {\bibfnamefont {S.}~\bibnamefont
  {Carruitero}}, \bibinfo {author} {\bibfnamefont {A.~C.}\ \bibnamefont
  {Duran}}, \bibinfo {author} {\bibfnamefont {G.}~\bibnamefont {Pisegna}},
  \bibinfo {author} {\bibfnamefont {M.~B.}\ \bibnamefont {Sturla}},\ and\
  \bibinfo {author} {\bibfnamefont {T.~S.}\ \bibnamefont {Grigera}},\
  }\bibfield  {title} {\bibinfo {title} {Inertial spin model of flocking with
  position-dependent forces},\ }\href
  {https://doi.org/10.1103/PhysRevE.110.014408} {\bibfield  {journal} {\bibinfo
   {journal} {Phys. Rev. E}\ }\textbf {\bibinfo {volume} {110}},\ \bibinfo
  {pages} {014408} (\bibinfo {year} {2024})}\BibitemShut {NoStop}%
\bibitem [{\citenamefont {{Gonz{\'a}lez-Albaladejo}}\ and\ \citenamefont
  {Bonilla}(2023)}]{gonzalez-albaladejoMeanfieldTheoryChaotic2023}%
  \BibitemOpen
  \bibfield  {author} {\bibinfo {author} {\bibfnamefont {R.}~\bibnamefont
  {{Gonz{\'a}lez-Albaladejo}}}\ and\ \bibinfo {author} {\bibfnamefont {L.~L.}\
  \bibnamefont {Bonilla}},\ }\bibfield  {title} {\bibinfo {title} {Mean-field
  theory of chaotic insect swarms},\ }\href
  {https://doi.org/10.1103/PhysRevE.107.L062601} {\bibfield  {journal}
  {\bibinfo  {journal} {Phys. Rev. E}\ }\textbf {\bibinfo {volume} {107}},\
  \bibinfo {pages} {L062601} (\bibinfo {year} {2023})}\BibitemShut {NoStop}%
\bibitem [{\citenamefont {D'Orsogna}\ \emph {et~al.}(2006)\citenamefont
  {D'Orsogna}, \citenamefont {Chuang}, \citenamefont {Bertozzi},\ and\
  \citenamefont {Chayes}}]{dorsognaSelfPropelledParticlesSoftCore2006}%
  \BibitemOpen
  \bibfield  {author} {\bibinfo {author} {\bibfnamefont {M.~R.}\ \bibnamefont
  {D'Orsogna}}, \bibinfo {author} {\bibfnamefont {Y.~L.}\ \bibnamefont
  {Chuang}}, \bibinfo {author} {\bibfnamefont {A.~L.}\ \bibnamefont
  {Bertozzi}},\ and\ \bibinfo {author} {\bibfnamefont {L.~S.}\ \bibnamefont
  {Chayes}},\ }\bibfield  {title} {\bibinfo {title} {Self-{{Propelled
  Particles}} with {{Soft-Core Interactions}}: {{Patterns}}, {{Stability}}, and
  {{Collapse}}},\ }\href {https://doi.org/10.1103/PhysRevLett.96.104302}
  {\bibfield  {journal} {\bibinfo  {journal} {Phys. Rev. Lett.}\ }\textbf
  {\bibinfo {volume} {96}},\ \bibinfo {pages} {104302} (\bibinfo {year}
  {2006})}\BibitemShut {NoStop}%
\bibitem [{\citenamefont {Ballerini}\ \emph {et~al.}(2008)\citenamefont
  {Ballerini}, \citenamefont {Cabibbo}, \citenamefont {Candelier},
  \citenamefont {Cavagna}, \citenamefont {Cisbani}, \citenamefont {Giardina},
  \citenamefont {Lecomte}, \citenamefont {Orlandi}, \citenamefont {Parisi},
  \citenamefont {Procaccini}, \citenamefont {Viale},\ and\ \citenamefont
  {Zdravkovic}}]{balleriniInteractionRulingAnimal2008}%
  \BibitemOpen
  \bibfield  {author} {\bibinfo {author} {\bibfnamefont {M.}~\bibnamefont
  {Ballerini}}, \bibinfo {author} {\bibfnamefont {N.}~\bibnamefont {Cabibbo}},
  \bibinfo {author} {\bibfnamefont {R.}~\bibnamefont {Candelier}}, \bibinfo
  {author} {\bibfnamefont {A.}~\bibnamefont {Cavagna}}, \bibinfo {author}
  {\bibfnamefont {E.}~\bibnamefont {Cisbani}}, \bibinfo {author} {\bibfnamefont
  {I.}~\bibnamefont {Giardina}}, \bibinfo {author} {\bibfnamefont
  {V.}~\bibnamefont {Lecomte}}, \bibinfo {author} {\bibfnamefont
  {A.}~\bibnamefont {Orlandi}}, \bibinfo {author} {\bibfnamefont
  {G.}~\bibnamefont {Parisi}}, \bibinfo {author} {\bibfnamefont
  {A.}~\bibnamefont {Procaccini}}, \bibinfo {author} {\bibfnamefont
  {M.}~\bibnamefont {Viale}},\ and\ \bibinfo {author} {\bibfnamefont
  {V.}~\bibnamefont {Zdravkovic}},\ }\bibfield  {title} {\bibinfo {title}
  {Interaction ruling animal collective behavior depends on topological rather
  than metric distance: {{Evidence}} from a field study},\ }\href
  {https://doi.org/10.1073/pnas.0711437105} {\bibfield  {journal} {\bibinfo
  {journal} {PNAS}\ }\textbf {\bibinfo {volume} {105}},\ \bibinfo {pages}
  {1232} (\bibinfo {year} {2008})}\BibitemShut {NoStop}%
\bibitem [{\citenamefont {Huth}\ and\ \citenamefont
  {Wissel}(1992)}]{huthSimulationMovementFish1992}%
  \BibitemOpen
  \bibfield  {author} {\bibinfo {author} {\bibfnamefont {A.}~\bibnamefont
  {Huth}}\ and\ \bibinfo {author} {\bibfnamefont {C.}~\bibnamefont {Wissel}},\
  }\bibfield  {title} {\bibinfo {title} {The simulation of the movement of fish
  schools},\ }\href {https://doi.org/10.1016/S0022-5193(05)80681-2} {\bibfield
  {journal} {\bibinfo  {journal} {J. Theor. Biol.}\ }\textbf {\bibinfo {volume}
  {156}},\ \bibinfo {pages} {365} (\bibinfo {year} {1992})}\BibitemShut
  {NoStop}%
\bibitem [{\citenamefont {Gorbonos}\ \emph {et~al.}(2016)\citenamefont
  {Gorbonos}, \citenamefont {Ianconescu}, \citenamefont {Puckett},
  \citenamefont {Ni}, \citenamefont {Ouellette},\ and\ \citenamefont
  {Gov}}]{gorbonosLongrangeAcousticInteractions2016}%
  \BibitemOpen
  \bibfield  {author} {\bibinfo {author} {\bibfnamefont {D.}~\bibnamefont
  {Gorbonos}}, \bibinfo {author} {\bibfnamefont {R.}~\bibnamefont
  {Ianconescu}}, \bibinfo {author} {\bibfnamefont {J.~G.}\ \bibnamefont
  {Puckett}}, \bibinfo {author} {\bibfnamefont {R.}~\bibnamefont {Ni}},
  \bibinfo {author} {\bibfnamefont {N.~T.}\ \bibnamefont {Ouellette}},\ and\
  \bibinfo {author} {\bibfnamefont {N.~S.}\ \bibnamefont {Gov}},\ }\bibfield
  {title} {\bibinfo {title} {Long-range acoustic interactions in insect swarms:
  An adaptive gravity model},\ }\href
  {https://doi.org/10.1088/1367-2630/18/7/073042} {\bibfield  {journal}
  {\bibinfo  {journal} {New J. Phys.}\ }\textbf {\bibinfo {volume} {18}},\
  \bibinfo {pages} {073042} (\bibinfo {year} {2016})}\BibitemShut {NoStop}%
\bibitem [{\citenamefont {Vicsek}\ and\ \citenamefont
  {Zafeiris}(2012)}]{vicsekCollectiveMotion2012}%
  \BibitemOpen
  \bibfield  {author} {\bibinfo {author} {\bibfnamefont {T.}~\bibnamefont
  {Vicsek}}\ and\ \bibinfo {author} {\bibfnamefont {A.}~\bibnamefont
  {Zafeiris}},\ }\bibfield  {title} {\bibinfo {title} {Collective motion},\
  }\href {https://doi.org/10.1016/j.physrep.2012.03.004} {\bibfield  {journal}
  {\bibinfo  {journal} {Physics Reports}\ }\bibinfo {series} {Collective
  Motion},\ \textbf {\bibinfo {volume} {517}},\ \bibinfo {pages} {71} (\bibinfo
  {year} {2012})}\BibitemShut {NoStop}%
\bibitem [{\citenamefont {Organization}(2018)}]{WHO2018}%
  \BibitemOpen
  \bibfield  {author} {\bibinfo {author} {\bibfnamefont {W.~H.}\ \bibnamefont
  {Organization}},\ }\href
  {https://iris.who.int/bitstream/handle/10665/272533/9789241514057-eng.pdfs}
  {\bibinfo {title} {Global report on insecticide resistance in malaria
  vectors: 2010–2016}} (\bibinfo {year} {2018})\BibitemShut {NoStop}%
\bibitem [{\citenamefont {Organization}(2020)}]{WHO2020}%
  \BibitemOpen
  \bibfield  {author} {\bibinfo {author} {\bibfnamefont {W.~H.}\ \bibnamefont
  {Organization}},\ }\href
  {https://www.who.int/news-room/fact-sheets/detail/vector-borne-diseases}
  {\bibinfo {title} {Vector-borne diseases}} (\bibinfo {year}
  {2020})\BibitemShut {NoStop}%
\bibitem [{\citenamefont {Messina}\ \emph {et~al.}(2015)\citenamefont
  {Messina}, \citenamefont {Brady}, \citenamefont {Pigott}, \citenamefont
  {Golding}, \citenamefont {Kraemer}, \citenamefont {Scott}, \citenamefont
  {Wint}, \citenamefont {Smith},\ and\ \citenamefont
  {Hay}}]{messinaManyProjectedFutures2015}%
  \BibitemOpen
  \bibfield  {author} {\bibinfo {author} {\bibfnamefont {J.~P.}\ \bibnamefont
  {Messina}}, \bibinfo {author} {\bibfnamefont {O.~J.}\ \bibnamefont {Brady}},
  \bibinfo {author} {\bibfnamefont {D.~M.}\ \bibnamefont {Pigott}}, \bibinfo
  {author} {\bibfnamefont {N.}~\bibnamefont {Golding}}, \bibinfo {author}
  {\bibfnamefont {M.~U.~G.}\ \bibnamefont {Kraemer}}, \bibinfo {author}
  {\bibfnamefont {T.~W.}\ \bibnamefont {Scott}}, \bibinfo {author}
  {\bibfnamefont {G.~R.~W.}\ \bibnamefont {Wint}}, \bibinfo {author}
  {\bibfnamefont {D.~L.}\ \bibnamefont {Smith}},\ and\ \bibinfo {author}
  {\bibfnamefont {S.~I.}\ \bibnamefont {Hay}},\ }\bibfield  {title} {\bibinfo
  {title} {The many projected futures of dengue},\ }\href
  {https://doi.org/10.1038/nrmicro3430} {\bibfield  {journal} {\bibinfo
  {journal} {Nat. Rev. Microbiol.}\ }\textbf {\bibinfo {volume} {13}},\
  \bibinfo {pages} {230} (\bibinfo {year} {2015})}\BibitemShut {NoStop}%
\bibitem [{\citenamefont {Su}\ \emph {et~al.}(2020)\citenamefont {Su},
  \citenamefont {Georgiades}, \citenamefont {Bagi}, \citenamefont {Kyrou},
  \citenamefont {Crisanti},\ and\ \citenamefont
  {Albert}}]{suAssessingAcousticBehaviour2020}%
  \BibitemOpen
  \bibfield  {author} {\bibinfo {author} {\bibfnamefont {M.~P.}\ \bibnamefont
  {Su}}, \bibinfo {author} {\bibfnamefont {M.}~\bibnamefont {Georgiades}},
  \bibinfo {author} {\bibfnamefont {J.}~\bibnamefont {Bagi}}, \bibinfo {author}
  {\bibfnamefont {K.}~\bibnamefont {Kyrou}}, \bibinfo {author} {\bibfnamefont
  {A.}~\bibnamefont {Crisanti}},\ and\ \bibinfo {author} {\bibfnamefont
  {J.~T.}\ \bibnamefont {Albert}},\ }\bibfield  {title} {\bibinfo {title}
  {Assessing the acoustic behaviour of {{Anopheles}} gambiae (s.l.) {{dsxF}}
  mutants: implications for vector control},\ }\href
  {https://doi.org/10.1186/s13071-020-04382-x} {\bibfield  {journal} {\bibinfo
  {journal} {Parasit. Vectors}\ }\textbf {\bibinfo {volume} {13}},\ \bibinfo
  {pages} {507} (\bibinfo {year} {2020})}\BibitemShut {NoStop}%
\bibitem [{\citenamefont {Andr{\'e}s}\ \emph {et~al.}(2020)\citenamefont
  {Andr{\'e}s}, \citenamefont {Su}, \citenamefont {Albert},\ and\ \citenamefont
  {Cator}}]{andresBuzzkillTargetingMosquito2020}%
  \BibitemOpen
  \bibfield  {author} {\bibinfo {author} {\bibfnamefont {M.}~\bibnamefont
  {Andr{\'e}s}}, \bibinfo {author} {\bibfnamefont {M.~P.}\ \bibnamefont {Su}},
  \bibinfo {author} {\bibfnamefont {J.}~\bibnamefont {Albert}},\ and\ \bibinfo
  {author} {\bibfnamefont {L.~J.}\ \bibnamefont {Cator}},\ }\bibfield  {title}
  {\bibinfo {title} {Buzzkill: Targeting the mosquito auditory system},\ }\href
  {https://doi.org/10.1016/j.cois.2020.04.003} {\bibfield  {journal} {\bibinfo
  {journal} {Curr. Opin. Insect Sci.}\ }\textbf {\bibinfo {volume} {40}},\
  \bibinfo {pages} {11} (\bibinfo {year} {2020})}\BibitemShut {NoStop}%
\bibitem [{\citenamefont {Seo}\ \emph {et~al.}(2019)\citenamefont {Seo},
  \citenamefont {Hedrick},\ and\ \citenamefont
  {Mittal}}]{seoMechanismScalingWing2019}%
  \BibitemOpen
  \bibfield  {author} {\bibinfo {author} {\bibfnamefont {J.-H.}\ \bibnamefont
  {Seo}}, \bibinfo {author} {\bibfnamefont {T.~L.}\ \bibnamefont {Hedrick}},\
  and\ \bibinfo {author} {\bibfnamefont {R.}~\bibnamefont {Mittal}},\
  }\bibfield  {title} {\bibinfo {title} {Mechanism and scaling of wing tone
  generation in mosquitoes},\ }\href {https://doi.org/10.1088/1748-3190/ab54fc}
  {\bibfield  {journal} {\bibinfo  {journal} {Bioinspir. Biomim.}\ }\textbf
  {\bibinfo {volume} {15}},\ \bibinfo {pages} {016008} (\bibinfo {year}
  {2019})}\BibitemShut {NoStop}%
\bibitem [{\citenamefont {Faber}\ \emph
  {et~al.}(2026{\natexlab{a}})\citenamefont {Faber}, \citenamefont
  {Alampounti}, \citenamefont {Georgiades}, \citenamefont {Albert},\ and\
  \citenamefont {Bozovic}}]{faberLocalizationFrequencyResolution2026}%
  \BibitemOpen
  \bibfield  {author} {\bibinfo {author} {\bibfnamefont {J.}~\bibnamefont
  {Faber}}, \bibinfo {author} {\bibfnamefont {A.~C.}\ \bibnamefont
  {Alampounti}}, \bibinfo {author} {\bibfnamefont {M.}~\bibnamefont
  {Georgiades}}, \bibinfo {author} {\bibfnamefont {J.~T.}\ \bibnamefont
  {Albert}},\ and\ \bibinfo {author} {\bibfnamefont {D.}~\bibnamefont
  {Bozovic}},\ }\bibfield  {title} {\bibinfo {title} {Localization and
  frequency resolution in antennal hearing of insects},\ }\href
  {https://doi.org/10.1103/7lt1-vzjr} {\bibfield  {journal} {\bibinfo
  {journal} {Phys. Rev. Res.}\ }\textbf {\bibinfo {volume} {8}},\ \bibinfo
  {pages} {013145} (\bibinfo {year} {2026}{\natexlab{a}})}\BibitemShut
  {NoStop}%
\bibitem [{\citenamefont {Sueur}\ \emph {et~al.}(2005)\citenamefont {Sueur},
  \citenamefont {Tuck},\ and\ \citenamefont
  {Robert}}]{sueurSoundRadiationFlying2005}%
  \BibitemOpen
  \bibfield  {author} {\bibinfo {author} {\bibfnamefont {J.}~\bibnamefont
  {Sueur}}, \bibinfo {author} {\bibfnamefont {E.~J.}\ \bibnamefont {Tuck}},\
  and\ \bibinfo {author} {\bibfnamefont {D.}~\bibnamefont {Robert}},\
  }\bibfield  {title} {\bibinfo {title} {Sound radiation around a flying fly},\
  }\href {https://doi.org/10.1121/1.1932227} {\bibfield  {journal} {\bibinfo
  {journal} {J. Acoust. Soc. Am.}\ }\textbf {\bibinfo {volume} {118}},\
  \bibinfo {pages} {530} (\bibinfo {year} {2005})}\BibitemShut {NoStop}%
\bibitem [{\citenamefont {Cribellier}\ \emph {et~al.}(2024)\citenamefont
  {Cribellier}, \citenamefont {Poda}, \citenamefont {Dabir{\'e}}, \citenamefont
  {Diabat{\'e}}, \citenamefont {Roux},\ and\ \citenamefont
  {Muijres}}]{cribellierComplexSwarmingDynamics2024}%
  \BibitemOpen
  \bibfield  {author} {\bibinfo {author} {\bibfnamefont {A.}~\bibnamefont
  {Cribellier}}, \bibinfo {author} {\bibfnamefont {B.~S.}\ \bibnamefont
  {Poda}}, \bibinfo {author} {\bibfnamefont {R.~K.}\ \bibnamefont
  {Dabir{\'e}}}, \bibinfo {author} {\bibfnamefont {A.}~\bibnamefont
  {Diabat{\'e}}}, \bibinfo {author} {\bibfnamefont {O.}~\bibnamefont {Roux}},\
  and\ \bibinfo {author} {\bibfnamefont {F.~T.}\ \bibnamefont {Muijres}},\
  }\bibfield  {title} {\bibinfo {title} {The complex swarming dynamics of
  malaria mosquitoes emerges from simple minimally-interactive behavioral
  rules},\ }\bibfield  {journal} {\bibinfo  {journal} {bioRxiv}\ }\href
  {https://doi.org/10.1101/2024.08.31.610631} {10.1101/2024.08.31.610631}
  (\bibinfo {year} {2024})\BibitemShut {NoStop}%
\bibitem [{\citenamefont {Puckett}\ \emph {et~al.}(2015)\citenamefont
  {Puckett}, \citenamefont {Ni},\ and\ \citenamefont
  {Ouellette}}]{puckettTimeFrequencyAnalysisReveals2015}%
  \BibitemOpen
  \bibfield  {author} {\bibinfo {author} {\bibfnamefont {J.~G.}\ \bibnamefont
  {Puckett}}, \bibinfo {author} {\bibfnamefont {R.}~\bibnamefont {Ni}},\ and\
  \bibinfo {author} {\bibfnamefont {N.~T.}\ \bibnamefont {Ouellette}},\
  }\bibfield  {title} {\bibinfo {title} {Time-{{Frequency Analysis Reveals
  Pairwise Interactions}} in {{Insect Swarms}}},\ }\href
  {https://doi.org/10.1103/PhysRevLett.114.258103} {\bibfield  {journal}
  {\bibinfo  {journal} {Phys. Rev. Lett.}\ }\textbf {\bibinfo {volume} {114}},\
  \bibinfo {pages} {258103} (\bibinfo {year} {2015})}\BibitemShut {NoStop}%
\bibitem [{\citenamefont {Poda}\ \emph {et~al.}(2024)\citenamefont {Poda},
  \citenamefont {Cribellier}, \citenamefont {Feug{\`e}re}, \citenamefont
  {Fatou}, \citenamefont {Nignan}, \citenamefont {Hien}, \citenamefont
  {M{\"u}ller}, \citenamefont {Gnankin{\'e}}, \citenamefont {Dabir{\'e}},
  \citenamefont {Diabat{\'e}}, \citenamefont {Muijres},\ and\ \citenamefont
  {Roux}}]{podaSpatialTemporalCharacteristics2024}%
  \BibitemOpen
  \bibfield  {author} {\bibinfo {author} {\bibfnamefont {B.~S.}\ \bibnamefont
  {Poda}}, \bibinfo {author} {\bibfnamefont {A.}~\bibnamefont {Cribellier}},
  \bibinfo {author} {\bibfnamefont {L.}~\bibnamefont {Feug{\`e}re}}, \bibinfo
  {author} {\bibfnamefont {M.}~\bibnamefont {Fatou}}, \bibinfo {author}
  {\bibfnamefont {C.}~\bibnamefont {Nignan}}, \bibinfo {author} {\bibfnamefont
  {D.~F. d.~S.}\ \bibnamefont {Hien}}, \bibinfo {author} {\bibfnamefont
  {P.}~\bibnamefont {M{\"u}ller}}, \bibinfo {author} {\bibfnamefont
  {O.}~\bibnamefont {Gnankin{\'e}}}, \bibinfo {author} {\bibfnamefont {R.~K.}\
  \bibnamefont {Dabir{\'e}}}, \bibinfo {author} {\bibfnamefont
  {A.}~\bibnamefont {Diabat{\'e}}}, \bibinfo {author} {\bibfnamefont {F.~T.}\
  \bibnamefont {Muijres}},\ and\ \bibinfo {author} {\bibfnamefont
  {O.}~\bibnamefont {Roux}},\ }\bibfield  {title} {\bibinfo {title} {Spatial
  and temporal characteristics of laboratory-induced {{Anopheles}} coluzzii
  swarms: {{Shape}}, structure, and flight kinematics},\ }\bibfield  {journal}
  {\bibinfo  {journal} {iScience}\ }\textbf {\bibinfo {volume} {27}},\ \href
  {https://doi.org/10.1016/j.isci.2024.111164} {10.1016/j.isci.2024.111164}
  (\bibinfo {year} {2024})\BibitemShut {NoStop}%
\bibitem [{\citenamefont
  {{Bennet-Clark}}(1998)}]{bennet-clarkSizeScaleEffects1998}%
  \BibitemOpen
  \bibfield  {author} {\bibinfo {author} {\bibfnamefont {H.~C.}\ \bibnamefont
  {{Bennet-Clark}}},\ }\bibfield  {title} {\bibinfo {title} {Size and {{Scale
  Effects}} as {{Constraints}} in {{Insect Sound Communication}}},\ }\href@noop
  {} {\bibfield  {journal} {\bibinfo  {journal} {Philos. Trans. R. Soc. B:
  Biol. Sci.}\ }\textbf {\bibinfo {volume} {353}},\ \bibinfo {pages} {407}
  (\bibinfo {year} {1998})}\BibitemShut {NoStop}%
\bibitem [{\citenamefont {G{\"o}pfert}\ and\ \citenamefont
  {Hennig}(2016)}]{gopfertHearingInsects2016}%
  \BibitemOpen
  \bibfield  {author} {\bibinfo {author} {\bibfnamefont {M.~C.}\ \bibnamefont
  {G{\"o}pfert}}\ and\ \bibinfo {author} {\bibfnamefont {R.~M.}\ \bibnamefont
  {Hennig}},\ }\bibfield  {title} {\bibinfo {title} {Hearing in {{Insects}}},\
  }\href {https://doi.org/10.1146/annurev-ento-010715-023631} {\bibfield
  {journal} {\bibinfo  {journal} {Ann. Rev. Entomol.}\ }\textbf {\bibinfo
  {volume} {61}},\ \bibinfo {pages} {257} (\bibinfo {year} {2016})}\BibitemShut
  {NoStop}%
\bibitem [{\citenamefont
  {R{\"o}mer}(2020)}]{romerDirectionalHearingInsects2020}%
  \BibitemOpen
  \bibfield  {author} {\bibinfo {author} {\bibfnamefont {H.}~\bibnamefont
  {R{\"o}mer}},\ }\bibfield  {title} {\bibinfo {title} {Directional hearing in
  insects: Biophysical, physiological and ecological challenges},\ }\href
  {https://doi.org/10.1242/jeb.203224} {\bibfield  {journal} {\bibinfo
  {journal} {J. Exp. Biol.}\ }\textbf {\bibinfo {volume} {223}},\ \bibinfo
  {pages} {jeb203224} (\bibinfo {year} {2020})}\BibitemShut {NoStop}%
\bibitem [{\citenamefont {Cavagna}\ \emph {et~al.}(2023)\citenamefont
  {Cavagna}, \citenamefont {Giardina}, \citenamefont {Gucciardino},
  \citenamefont {Iacomelli}, \citenamefont {Lombardi}, \citenamefont {Melillo},
  \citenamefont {Monacchia}, \citenamefont {Parisi}, \citenamefont {Peirce},\
  and\ \citenamefont {Spaccapelo}}]{cavagnaCharacterizationLabbasedSwarms2023}%
  \BibitemOpen
  \bibfield  {author} {\bibinfo {author} {\bibfnamefont {A.}~\bibnamefont
  {Cavagna}}, \bibinfo {author} {\bibfnamefont {I.}~\bibnamefont {Giardina}},
  \bibinfo {author} {\bibfnamefont {M.~A.}\ \bibnamefont {Gucciardino}},
  \bibinfo {author} {\bibfnamefont {G.}~\bibnamefont {Iacomelli}}, \bibinfo
  {author} {\bibfnamefont {M.}~\bibnamefont {Lombardi}}, \bibinfo {author}
  {\bibfnamefont {S.}~\bibnamefont {Melillo}}, \bibinfo {author} {\bibfnamefont
  {G.}~\bibnamefont {Monacchia}}, \bibinfo {author} {\bibfnamefont
  {L.}~\bibnamefont {Parisi}}, \bibinfo {author} {\bibfnamefont {M.~J.}\
  \bibnamefont {Peirce}},\ and\ \bibinfo {author} {\bibfnamefont
  {R.}~\bibnamefont {Spaccapelo}},\ }\bibfield  {title} {\bibinfo {title}
  {Characterization of lab-based swarms of {{Anopheles}} gambiae mosquitoes
  using {{3D-video}} tracking},\ }\href
  {https://doi.org/10.1038/s41598-023-34842-0} {\bibfield  {journal} {\bibinfo
  {journal} {Sci. Rep.}\ }\textbf {\bibinfo {volume} {13}},\ \bibinfo {pages}
  {8745} (\bibinfo {year} {2023})}\BibitemShut {NoStop}%
\bibitem [{\citenamefont {Gorbonos}\ \emph {et~al.}(2020)\citenamefont
  {Gorbonos}, \citenamefont {Puckett}, \citenamefont {{van der Vaart}},
  \citenamefont {Sinhuber}, \citenamefont {Ouellette},\ and\ \citenamefont
  {Gov}}]{gorbonosPairFormationInsect2020}%
  \BibitemOpen
  \bibfield  {author} {\bibinfo {author} {\bibfnamefont {D.}~\bibnamefont
  {Gorbonos}}, \bibinfo {author} {\bibfnamefont {J.~G.}\ \bibnamefont
  {Puckett}}, \bibinfo {author} {\bibfnamefont {K.}~\bibnamefont {{van der
  Vaart}}}, \bibinfo {author} {\bibfnamefont {M.}~\bibnamefont {Sinhuber}},
  \bibinfo {author} {\bibfnamefont {N.~T.}\ \bibnamefont {Ouellette}},\ and\
  \bibinfo {author} {\bibfnamefont {N.~S.}\ \bibnamefont {Gov}},\ }\bibfield
  {title} {\bibinfo {title} {Pair formation in insect swarms driven by adaptive
  long-range interactions},\ }\href {https://doi.org/10.1098/rsif.2020.0367}
  {\bibfield  {journal} {\bibinfo  {journal} {J. Roy. Soc. Int.}\ }\textbf
  {\bibinfo {volume} {17}},\ \bibinfo {pages} {20200367} (\bibinfo {year}
  {2020})}\BibitemShut {NoStop}%
\bibitem [{\citenamefont {Kelley}\ and\ \citenamefont
  {Ouellette}(2013)}]{kelleyEmergentDynamicsLaboratory2013}%
  \BibitemOpen
  \bibfield  {author} {\bibinfo {author} {\bibfnamefont {D.~H.}\ \bibnamefont
  {Kelley}}\ and\ \bibinfo {author} {\bibfnamefont {N.~T.}\ \bibnamefont
  {Ouellette}},\ }\bibfield  {title} {\bibinfo {title} {Emergent dynamics of
  laboratory insect swarms},\ }\href {https://doi.org/10.1038/srep01073}
  {\bibfield  {journal} {\bibinfo  {journal} {Sci. Rep.}\ }\textbf {\bibinfo
  {volume} {3}},\ \bibinfo {pages} {1073} (\bibinfo {year} {2013})}\BibitemShut
  {NoStop}%
\bibitem [{\citenamefont {Wellens}\ \emph {et~al.}(2003)\citenamefont
  {Wellens}, \citenamefont {Shatokhin},\ and\ \citenamefont
  {Buchleitner}}]{wellensStochasticResonance2003}%
  \BibitemOpen
  \bibfield  {author} {\bibinfo {author} {\bibfnamefont {T.}~\bibnamefont
  {Wellens}}, \bibinfo {author} {\bibfnamefont {V.}~\bibnamefont {Shatokhin}},\
  and\ \bibinfo {author} {\bibfnamefont {A.}~\bibnamefont {Buchleitner}},\
  }\bibfield  {title} {\bibinfo {title} {Stochastic resonance},\ }\href
  {https://doi.org/10.1088/0034-4885/67/1/R02} {\bibfield  {journal} {\bibinfo
  {journal} {Rep. Prog. Phys.}\ }\textbf {\bibinfo {volume} {67}},\ \bibinfo
  {pages} {45} (\bibinfo {year} {2003})}\BibitemShut {NoStop}%
\bibitem [{\citenamefont
  {Reynolds}(2023)}]{reynoldsStochasticityMayGenerate2023}%
  \BibitemOpen
  \bibfield  {author} {\bibinfo {author} {\bibfnamefont {A.~M.}\ \bibnamefont
  {Reynolds}},\ }\bibfield  {title} {\bibinfo {title} {Stochasticity may
  generate coherent motion in bird flocks},\ }\href
  {https://doi.org/10.1088/1478-3975/acbad7} {\bibfield  {journal} {\bibinfo
  {journal} {Phys. Biol.}\ }\textbf {\bibinfo {volume} {20}},\ \bibinfo {pages}
  {025002} (\bibinfo {year} {2023})}\BibitemShut {NoStop}%
\bibitem [{\citenamefont {Arthur}\ \emph {et~al.}(2014)\citenamefont {Arthur},
  \citenamefont {Emr}, \citenamefont {Wyttenbach},\ and\ \citenamefont
  {Hoy}}]{arthurMosquitoAedesAegypti2014}%
  \BibitemOpen
  \bibfield  {author} {\bibinfo {author} {\bibfnamefont {B.~J.}\ \bibnamefont
  {Arthur}}, \bibinfo {author} {\bibfnamefont {K.~S.}\ \bibnamefont {Emr}},
  \bibinfo {author} {\bibfnamefont {R.~A.}\ \bibnamefont {Wyttenbach}},\ and\
  \bibinfo {author} {\bibfnamefont {R.~R.}\ \bibnamefont {Hoy}},\ }\bibfield
  {title} {\bibinfo {title} {Mosquito ( {{Aedes}} aegypti ) flight tones:
  {{Frequency}}, harmonicity, spherical spreading, and phase relationships},\
  }\href {https://doi.org/10.1121/1.4861233} {\bibfield  {journal} {\bibinfo
  {journal} {J. Acoust. Soc. Am.}\ }\textbf {\bibinfo {volume} {135}},\
  \bibinfo {pages} {933} (\bibinfo {year} {2014})}\BibitemShut {NoStop}%
\bibitem [{\citenamefont {Faber}\ \emph
  {et~al.}(2026{\natexlab{b}})\citenamefont {Faber}, \citenamefont
  {Alampounti}, \citenamefont {Georgiades}, \citenamefont {Albert},\ and\
  \citenamefont {Bozovic}}]{faberSoundMaskingStrategies2026}%
  \BibitemOpen
  \bibfield  {author} {\bibinfo {author} {\bibfnamefont {J.}~\bibnamefont
  {Faber}}, \bibinfo {author} {\bibfnamefont {A.~C.}\ \bibnamefont
  {Alampounti}}, \bibinfo {author} {\bibfnamefont {M.}~\bibnamefont
  {Georgiades}}, \bibinfo {author} {\bibfnamefont {J.~T.}\ \bibnamefont
  {Albert}},\ and\ \bibinfo {author} {\bibfnamefont {D.}~\bibnamefont
  {Bozovic}},\ }\bibfield  {title} {\bibinfo {title} {Sound masking strategies
  for interference with mosquito hearing},\ }\href
  {https://doi.org/10.1063/5.0340103} {\bibfield  {journal} {\bibinfo
  {journal} {Chaos: An Interdisciplinary Journal of Nonlinear Science}\
  }\textbf {\bibinfo {volume} {36}},\ \bibinfo {pages} {073137} (\bibinfo
  {year} {2026}{\natexlab{b}})}\BibitemShut {NoStop}%
\bibitem [{\citenamefont {Gorbonos}\ and\ \citenamefont
  {Gov}(2017)}]{gorbonosStableSwarmingUsing2017}%
  \BibitemOpen
  \bibfield  {author} {\bibinfo {author} {\bibfnamefont {D.}~\bibnamefont
  {Gorbonos}}\ and\ \bibinfo {author} {\bibfnamefont {N.~S.}\ \bibnamefont
  {Gov}},\ }\bibfield  {title} {\bibinfo {title} {Stable swarming using
  adaptive long-range interactions},\ }\href
  {https://doi.org/10.1103/PhysRevE.95.042405} {\bibfield  {journal} {\bibinfo
  {journal} {Phys. Rev. E}\ }\textbf {\bibinfo {volume} {95}},\ \bibinfo
  {pages} {042405} (\bibinfo {year} {2017})}\BibitemShut {NoStop}%
\bibitem [{\citenamefont {Gupta}\ \emph {et~al.}(2024)\citenamefont {Gupta},
  \citenamefont {Cribellier}, \citenamefont {Poda}, \citenamefont {Roux},
  \citenamefont {Muijres},\ and\ \citenamefont
  {Riffell}}]{guptaMosquitoesIntegrateVisual2024}%
  \BibitemOpen
  \bibfield  {author} {\bibinfo {author} {\bibfnamefont {S.}~\bibnamefont
  {Gupta}}, \bibinfo {author} {\bibfnamefont {A.}~\bibnamefont {Cribellier}},
  \bibinfo {author} {\bibfnamefont {S.~B.}\ \bibnamefont {Poda}}, \bibinfo
  {author} {\bibfnamefont {O.}~\bibnamefont {Roux}}, \bibinfo {author}
  {\bibfnamefont {F.~T.}\ \bibnamefont {Muijres}},\ and\ \bibinfo {author}
  {\bibfnamefont {J.~A.}\ \bibnamefont {Riffell}},\ }\bibfield  {title}
  {\bibinfo {title} {Mosquitoes integrate visual and acoustic cues to mediate
  conspecific interactions in swarms},\ }\href
  {https://doi.org/10.1016/j.cub.2024.07.043} {\bibfield  {journal} {\bibinfo
  {journal} {Curr. Biol.}\ }\textbf {\bibinfo {volume} {34}},\ \bibinfo {pages}
  {4091} (\bibinfo {year} {2024})}\BibitemShut {NoStop}%
\bibitem [{\citenamefont {Poda}\ \emph {et~al.}(2022)\citenamefont {Poda},
  \citenamefont {Buatois}, \citenamefont {Lapeyre}, \citenamefont {Dormont},
  \citenamefont {Diabat{\'e}}, \citenamefont {Gnankin{\'e}}, \citenamefont
  {Dabir{\'e}},\ and\ \citenamefont {Roux}}]{podaNoEvidenceLongrange2022}%
  \BibitemOpen
  \bibfield  {author} {\bibinfo {author} {\bibfnamefont {S.~B.}\ \bibnamefont
  {Poda}}, \bibinfo {author} {\bibfnamefont {B.}~\bibnamefont {Buatois}},
  \bibinfo {author} {\bibfnamefont {B.}~\bibnamefont {Lapeyre}}, \bibinfo
  {author} {\bibfnamefont {L.}~\bibnamefont {Dormont}}, \bibinfo {author}
  {\bibfnamefont {A.}~\bibnamefont {Diabat{\'e}}}, \bibinfo {author}
  {\bibfnamefont {O.}~\bibnamefont {Gnankin{\'e}}}, \bibinfo {author}
  {\bibfnamefont {R.~K.}\ \bibnamefont {Dabir{\'e}}},\ and\ \bibinfo {author}
  {\bibfnamefont {O.}~\bibnamefont {Roux}},\ }\bibfield  {title} {\bibinfo
  {title} {No evidence for long-range male sex pheromones in two malaria
  mosquitoes},\ }\href {https://doi.org/10.1038/s41559-022-01869-x} {\bibfield
  {journal} {\bibinfo  {journal} {Nat. Ecol. Evol.}\ }\textbf {\bibinfo
  {volume} {6}},\ \bibinfo {pages} {1676} (\bibinfo {year} {2022})}\BibitemShut
  {NoStop}%
\bibitem [{\citenamefont {Faber}\ \emph {et~al.}(2025)\citenamefont {Faber},
  \citenamefont {Alampounti}, \citenamefont {Georgiades}, \citenamefont
  {Albert},\ and\ \citenamefont
  {Bozovic}}]{faberMosquitoinspiredTheoreticalFramework2025a}%
  \BibitemOpen
  \bibfield  {author} {\bibinfo {author} {\bibfnamefont {J.}~\bibnamefont
  {Faber}}, \bibinfo {author} {\bibfnamefont {A.~C.}\ \bibnamefont
  {Alampounti}}, \bibinfo {author} {\bibfnamefont {M.}~\bibnamefont
  {Georgiades}}, \bibinfo {author} {\bibfnamefont {J.~T.}\ \bibnamefont
  {Albert}},\ and\ \bibinfo {author} {\bibfnamefont {D.}~\bibnamefont
  {Bozovic}},\ }\bibfield  {title} {\bibinfo {title} {A mosquito-inspired
  theoretical framework for acoustic signal detection},\ }\href
  {https://doi.org/10.1073/pnas.2500938122} {\bibfield  {journal} {\bibinfo
  {journal} {PNAS}\ }\textbf {\bibinfo {volume} {122}},\ \bibinfo {pages}
  {e2500938122} (\bibinfo {year} {2025})}\BibitemShut {NoStop}%
\bibitem [{\citenamefont {Aldersley}\ \emph {et~al.}(2017)\citenamefont
  {Aldersley}, \citenamefont {Champneys}, \citenamefont {Homer}, \citenamefont
  {Bode},\ and\ \citenamefont {Robert}}]{aldersleyEmergentAcousticOrder2017}%
  \BibitemOpen
  \bibfield  {author} {\bibinfo {author} {\bibfnamefont {A.}~\bibnamefont
  {Aldersley}}, \bibinfo {author} {\bibfnamefont {A.}~\bibnamefont
  {Champneys}}, \bibinfo {author} {\bibfnamefont {M.}~\bibnamefont {Homer}},
  \bibinfo {author} {\bibfnamefont {N.~W.~F.}\ \bibnamefont {Bode}},\ and\
  \bibinfo {author} {\bibfnamefont {D.}~\bibnamefont {Robert}},\ }\bibfield
  {title} {\bibinfo {title} {Emergent acoustic order in arrays of mosquitoes},\
  }\href {https://doi.org/10.1016/j.cub.2017.09.055} {\bibfield  {journal}
  {\bibinfo  {journal} {Curr. Biol.}\ }\textbf {\bibinfo {volume} {27}},\
  \bibinfo {pages} {R1208} (\bibinfo {year} {2017})}\BibitemShut {NoStop}%
\bibitem [{\citenamefont {Iacomelli}\ \emph {et~al.}(2026)\citenamefont
  {Iacomelli}, \citenamefont {Lombardi}, \citenamefont {Parisi}, \citenamefont
  {Fiorini}, \citenamefont {Grucci}, \citenamefont {Lavorgna}, \citenamefont
  {Ligato}, \citenamefont {Zarcone}, \citenamefont {Peirce}, \citenamefont
  {Melillo},\ and\ \citenamefont
  {Spaccapelo}}]{iacomelliExperimentalEvidenceMale2026}%
  \BibitemOpen
  \bibfield  {author} {\bibinfo {author} {\bibfnamefont {G.}~\bibnamefont
  {Iacomelli}}, \bibinfo {author} {\bibfnamefont {M.}~\bibnamefont {Lombardi}},
  \bibinfo {author} {\bibfnamefont {L.}~\bibnamefont {Parisi}}, \bibinfo
  {author} {\bibfnamefont {M.}~\bibnamefont {Fiorini}}, \bibinfo {author}
  {\bibfnamefont {F.}~\bibnamefont {Grucci}}, \bibinfo {author} {\bibfnamefont
  {A.}~\bibnamefont {Lavorgna}}, \bibinfo {author} {\bibfnamefont
  {M.}~\bibnamefont {Ligato}}, \bibinfo {author} {\bibfnamefont
  {G.}~\bibnamefont {Zarcone}}, \bibinfo {author} {\bibfnamefont {M.~J.}\
  \bibnamefont {Peirce}}, \bibinfo {author} {\bibfnamefont {S.}~\bibnamefont
  {Melillo}},\ and\ \bibinfo {author} {\bibfnamefont {R.}~\bibnamefont
  {Spaccapelo}},\ }\bibfield  {title} {\bibinfo {title} {Experimental evidence
  of male--male interaction in laboratory swarms of anopheles gambiae
  mosquitoes},\ }\bibfield  {journal} {\bibinfo  {journal} {Sci. Rep.}\
  }\textbf {\bibinfo {volume} {(to be published)}},\ \href
  {https://doi.org/10.1038/s41598-026-59111-8} {10.1038/s41598-026-59111-8}
  (\bibinfo {year} {2026})\BibitemShut {NoStop}%
\bibitem [{\citenamefont {Puckett}\ \emph {et~al.}(2014)\citenamefont
  {Puckett}, \citenamefont {Kelley},\ and\ \citenamefont
  {Ouellette}}]{puckettSearchingEffectiveForces2014}%
  \BibitemOpen
  \bibfield  {author} {\bibinfo {author} {\bibfnamefont {J.~G.}\ \bibnamefont
  {Puckett}}, \bibinfo {author} {\bibfnamefont {D.~H.}\ \bibnamefont
  {Kelley}},\ and\ \bibinfo {author} {\bibfnamefont {N.~T.}\ \bibnamefont
  {Ouellette}},\ }\bibfield  {title} {\bibinfo {title} {Searching for effective
  forces in laboratory insect swarms},\ }\href
  {https://doi.org/10.1038/srep04766} {\bibfield  {journal} {\bibinfo
  {journal} {Sci. Rep.}\ }\textbf {\bibinfo {volume} {4}},\ \bibinfo {pages}
  {4766} (\bibinfo {year} {2014})}\BibitemShut {NoStop}%
\bibitem [{\citenamefont {O'Byrne}\ and\ \citenamefont
  {Jerbi}(2022)}]{obyrneHowCriticalBrain2022}%
  \BibitemOpen
  \bibfield  {author} {\bibinfo {author} {\bibfnamefont {J.}~\bibnamefont
  {O'Byrne}}\ and\ \bibinfo {author} {\bibfnamefont {K.}~\bibnamefont
  {Jerbi}},\ }\bibfield  {title} {\bibinfo {title} {How critical is brain
  criticality?},\ }\href {https://doi.org/10.1016/j.tins.2022.08.007}
  {\bibfield  {journal} {\bibinfo  {journal} {Trends Neurosci.}\ }\textbf
  {\bibinfo {volume} {45}},\ \bibinfo {pages} {820} (\bibinfo {year}
  {2022})}\BibitemShut {NoStop}%
\bibitem [{\citenamefont {Paschen}\ and\ \citenamefont
  {Si}(2021)}]{paschenQuantumPhasesDriven2021}%
  \BibitemOpen
  \bibfield  {author} {\bibinfo {author} {\bibfnamefont {S.}~\bibnamefont
  {Paschen}}\ and\ \bibinfo {author} {\bibfnamefont {Q.}~\bibnamefont {Si}},\
  }\bibfield  {title} {\bibinfo {title} {Quantum phases driven by strong
  correlations},\ }\href {https://doi.org/10.1038/s42254-020-00262-6}
  {\bibfield  {journal} {\bibinfo  {journal} {Nat. Rev. Phys.}\ }\textbf
  {\bibinfo {volume} {3}},\ \bibinfo {pages} {9} (\bibinfo {year}
  {2021})}\BibitemShut {NoStop}%
\bibitem [{\citenamefont {Bahri}\ \emph {et~al.}(2020)\citenamefont {Bahri},
  \citenamefont {Kadmon}, \citenamefont {Pennington}, \citenamefont
  {Schoenholz}, \citenamefont {{Sohl-Dickstein}},\ and\ \citenamefont
  {Ganguli}}]{bahriStatisticalMechanicsDeep2020}%
  \BibitemOpen
  \bibfield  {author} {\bibinfo {author} {\bibfnamefont {Y.}~\bibnamefont
  {Bahri}}, \bibinfo {author} {\bibfnamefont {J.}~\bibnamefont {Kadmon}},
  \bibinfo {author} {\bibfnamefont {J.}~\bibnamefont {Pennington}}, \bibinfo
  {author} {\bibfnamefont {S.~S.}\ \bibnamefont {Schoenholz}}, \bibinfo
  {author} {\bibfnamefont {J.}~\bibnamefont {{Sohl-Dickstein}}},\ and\ \bibinfo
  {author} {\bibfnamefont {S.}~\bibnamefont {Ganguli}},\ }\bibfield  {title}
  {\bibinfo {title} {Statistical {{Mechanics}} of {{Deep Learning}}},\ }\href
  {https://doi.org/10.1146/annurev-conmatphys-031119-050745} {\bibfield
  {journal} {\bibinfo  {journal} {Annu. Rev. Condens. Matter Phys.}\ }\textbf
  {\bibinfo {volume} {11}},\ \bibinfo {pages} {501} (\bibinfo {year}
  {2020})}\BibitemShut {NoStop}%
\bibitem [{\citenamefont {Watkins}\ \emph {et~al.}(2016)\citenamefont
  {Watkins}, \citenamefont {Pruessner}, \citenamefont {Chapman}, \citenamefont
  {Crosby},\ and\ \citenamefont {Jensen}}]{watkins25YearsSelforganized2016}%
  \BibitemOpen
  \bibfield  {author} {\bibinfo {author} {\bibfnamefont {N.~W.}\ \bibnamefont
  {Watkins}}, \bibinfo {author} {\bibfnamefont {G.}~\bibnamefont {Pruessner}},
  \bibinfo {author} {\bibfnamefont {S.~C.}\ \bibnamefont {Chapman}}, \bibinfo
  {author} {\bibfnamefont {N.~B.}\ \bibnamefont {Crosby}},\ and\ \bibinfo
  {author} {\bibfnamefont {H.~J.}\ \bibnamefont {Jensen}},\ }\bibfield  {title}
  {\bibinfo {title} {25 {{Years}} of {{Self-organized Criticality}}:
  {{Concepts}} and {{Controversies}}},\ }\href
  {https://doi.org/10.1007/s11214-015-0155-x} {\bibfield  {journal} {\bibinfo
  {journal} {Space Sci. Rev.}\ }\textbf {\bibinfo {volume} {198}},\ \bibinfo
  {pages} {3} (\bibinfo {year} {2016})}\BibitemShut {NoStop}%
\bibitem [{\citenamefont {Faber}\ \emph
  {et~al.}(2026{\natexlab{c}})\citenamefont {Faber}, \citenamefont {Boots},\
  and\ \citenamefont {Bozovic}}]{DataAvailability}%
  \BibitemOpen
  \bibfield  {author} {\bibinfo {author} {\bibfnamefont {J.}~\bibnamefont
  {Faber}}, \bibinfo {author} {\bibfnamefont {A.}~\bibnamefont {Boots}},\ and\
  \bibinfo {author} {\bibfnamefont {D.}~\bibnamefont {Bozovic}},\ }\href@noop
  {} {\bibinfo {title} {Python code for all analysis and figure generation}},\
  \bibinfo {howpublished}
  {\href{https://github.com/jfaber3/code-for-Mosquito-swarms-operate-at-the-edge-of-instability.git}{Available
  on GitHub}} (\bibinfo {year} {2026}{\natexlab{c}})\BibitemShut {NoStop}%
\end{thebibliography}%

\section*{End Matter}

\textit{Length scale of swarm}---Consider a single insect orbiting a swarm that is concentrated at one point and stationary. For simplicity, we consider a circular orbit with radius, $R_s$. The circumference of the circle that the trajectory follows is $2\pi R_s \approx n\ell$, where $n$ is the number of time steps required for the insect to complete one full orbit. Using the Eq. \ref{eq:theta_diff} with no noise, the orientation angle changes by $\pi R_s /L$ every time step. After $n$ time steps, the orientation will complete a $2\pi$ rotation. Therefore, $n\pi R_s /L = 2\pi$. Combining these two equations by eliminating $n$, we can solve for the radius of this trajectory:

\begin{eqnarray}
R_s = \sqrt{\frac{\ell L}{\pi}}
\end{eqnarray}

\noindent  In practice, the swarms are actually several times larger than this, since circular orbits are unlikely. However, this gives us an upper bound on the curvature of the trajectories.

\textit{Pair formation dynamics}---Consider the case of just two insects, traveling along the positive $x-$axis. Let their initial orientations be parallel to the $x-$axis and their initial positions be $(0, \pm y(t=0))$. We will use $y(t)$ to represent the first insect's distance from the $x-$axis and $\theta(t)$ to represent the angle its orientation makes with the $x-$axis. By symmetry, the position and orientation of the second insect are $-y(t)$, and $-\theta(t)$, respectively. Using Eq. \ref{eq:mv_dot}, the differential equation for the first insect is

\begin{eqnarray}
\frac{m v_0}{k}\frac{d\theta_1}{dt} =  D_{1, y}\cos\theta_1 - D_{1, x}\sin\theta_1,
\end{eqnarray}

\noindent where $D_{1,x} = 0$ due to the configuration of the problem and $D_{1,y} = -2y(t)$ is the distance between the two insects. Our equation then becomes

\begin{eqnarray}
\frac{m v_0}{k}\frac{d\theta}{dt} =  -2y\cos\theta,
\label{eq:int_theta_dot}
\end{eqnarray}

\noindent where we have dropped the subscripts for simplicity. We invoke the constant speed constraint $\dot{y}(t) = v_0\sin\theta(t)$, which has derivative $\ddot{y}(t) = v_0\cos\theta \frac{d\theta}{dt}$. Combining this with Eq. \ref{eq:int_theta_dot}, we find a second-order differential equation of a single variable:

\begin{eqnarray} \label{eq:y_ddot}
\ddot{y}  + \omega^2_0 y (1-\frac{\dot{y}^2}{v_0^2}) = 0,
\end{eqnarray}

\noindent where we have defined $\omega_0 = \sqrt{\frac{2k}{m}}$, to be the characteristic frequency of the system. At small-amplitude orbits, we should expect this system to behave like a harmonic oscillator and the trajectories to be sinusoidal. However, for larger amplitude oscillations, our constant speed constraint distorts the trajectories.

\noindent At the only fixed point in the system, the Jacobian has eigenvalues $\pm i \omega_0$. So, $\omega_0$ is, therefore, the characteristic frequency of pair orbits. The purely imaginary eigenvalues indicate that this is a center manifold (center point), just like the simple harmonic oscillator. This means that the radial direction of the orbits is neutrally stable and that steady-state oscillations of various amplitude may occur.

By incorporating the initial conditions, we can also express Eq. \ref{eq:y_ddot} as a first-order differential equation of just one variable: 

\begin{eqnarray}  \label{eq:y_dot}
\dot{y}^2 = v_0^2 + \big(\dot{y}(0)^2 - v_0^2\big) e^{\frac{\omega_0^2}{v_0^2} \big[y^2 - y(0)^2\big]}.
\end{eqnarray}

\noindent Lastly, we can use Eq. \ref{eq:y_dot} to find the relationship between the amplitude of pair orbits and the maximum angle between orientations of the insects. Since the maximum angle occurs at $y=0$, we use $y(0) = 0$ and $\dot{y}(0) = v_0 \sin\theta_{max}$. Using these initial conditions and setting $\dot{y} = 0$ to find the turning point, we arrive at 

\begin{eqnarray}  \label{eq:y_max}
\cos{^2 \theta_{max}} = e^{-(\frac{\omega_0}{v_0})^2 y_{max}^2}
\end{eqnarray}

\noindent which will be helpful later for estimating stability conditions.

\textit{Pair stability near a swarm}---We now consider the same pair of insects flying away from a swarm of $N$ other insects. We want to know how tight the pair's orbit must be in order to escape the swarm. In other words, we want to know at what point, one of the individuals' attraction to the other pair member is equal to its attraction to the swarm. To do this, we once again invoke Eq. \ref{eq:theta_diff}. The weighted mean field vector takes the form,

\begin{eqnarray} 
\vec{D} = \frac{-2y\hat{y}/(2y)^\gamma - Nx\hat{x}/x^\gamma}{1/(2y)^\gamma + N/x^\gamma},
\end{eqnarray}

\noindent where we are assuming the swarm is concentrated at the origin. $x(t)$ is the distance from the pair center to the swarm and $2y(t)$ is the distance between the two pair members. From Eq. \ref{eq:theta_diff}, we see that breaking point where $\dot{\theta} = 0$ occurs at the angle where $\tan\theta = D_y/D_x$. This allows us to estimate the critical angle,

\begin{eqnarray}  \label{eq:crit_angle}
\tan\theta_c = \frac{1}{N} \bigg( \frac{x(t)}{2y(t)}\bigg)^{\gamma-1},
\end{eqnarray}

\noindent at which a pair will break, due to attraction to the swarm.

\textit{Analytic calculation of lower bifurcation curve}---We now estimate the bifurcation curve between pair stability and swarm stability. The calculation is similar to those performed in the kinetic theory of gasses. We will first compute the pair-formation rate ($\lambda_f$), which tells us, on average, how many pair bonds form per unit time. We will then compute the expected lifetime of these pairs and thus the annihilation rate ($\lambda_a$). Finally, we will find the critical noise strength ($\eta_c$) at which these two rates are equal.

We first calculate the expected relative speed, $\langle |\vec{v}_{rel}| \rangle$, between two insects passing by each other. We my express the square of the relative speed as

\begin{eqnarray} 
|\vec{v}_{rel}|^2  = 2v_0^2 - 2v_0^2\cos(\Delta\phi) = 4v_0^2\sin{^2\Big(\frac{\Delta\phi}{2}\Big)},
\end{eqnarray}

\noindent where $\Delta\phi$ is the angle between the directions of travel of the two insects. We must integrate $|\vec{v}_{rel}|$ over all angles, keeping only the angles where pairs are likely to form. We use Use Eq. \ref{eq:crit_angle} to estimate the maximum angle of passage where a pair may still form:

\begin{eqnarray} 
&\langle|\vec{v}_{rel}|  \mathbf{1}_{|\Delta\phi| < 2\theta_c} \rangle = \frac{1}{2\pi} \int_{-2\theta_c}^{2\theta_c} 2v_0|\sin(\frac{\Delta\phi}{2})| d\Delta\phi \notag     \\  \notag \\
    &= \frac{4v_0}{\pi}\Big(1 - \frac{1}{\sqrt{1 + q^2}} \Big ) 
    \approx \frac{4v_0}{\pi}\Big( \frac{q^2}{2} - \frac{3q^4}{8} + \frac{5q^6}{16} \Big), \notag \\ \\ 
&q = \frac{1}{N}\big(\frac{r}{a}\big)^{\gamma-1} \notag 
\end{eqnarray}

\noindent where $\mathbf{1}$ is the indicator function, $r$ is the distance of our reference insect from the swarm center and $a$ is the separation threshold for which a pair may form.

The number of pairs formed around our reference insect within time interval $dt$ are contained in the area element $dA = 2a \times \langle|\vec{v}_{rel}|  \mathbf{1}_{|\Delta\phi| < 2\theta_c} \rangle dt$, where $a$ is the interaction radius, or the maximum range at which pairs may form. We estimate the interaction radius using Eq. \ref{eq:y_max}. For $\theta_{max} = \pi/4$ (which corresponds to a $\pi/2$ angle between the pair), we find $a = 2y_{max} = 2\frac{v_0}{\omega_0}\sqrt{\log(2)} = R_s\sqrt{2\log(2)} \approx 1.18 R_s$.

With these conditions, we may now set up the integral over space to compute the pair formation rate,

\begin{eqnarray} 
\lambda_f = \frac{1}{4} \int_0^{2\pi}\int_0^\infty \rho(r)^2 \bigg[\frac{8v_0 a}{\pi}\Big(\frac{q^2}{2} - \frac{3q^4}{8} + \frac{5q^6}{16}\Big)\bigg] r dr d\theta, \notag
\end{eqnarray}

\noindent where $\rho(r) = \frac{N}{2\pi r_0^2} e^{-\frac{r^2}{2r_0^2}}$ is the swarm density. Since we are counting pairs, we integrate over the square of the density and include a factor of $\frac{1}{2}$ to avoid double counting. An additional factor of $\frac{1}{2}$ appears to remove transient insect pairs that are headed in the direction of the swarm, rather than away from it. Evaluating the integrals, we find

\begin{eqnarray} 
\lambda_f &= \frac{v_0}{4\pi^2 r_0}  \bigg[\Big(\frac{r_0}{a}\Big)^{2\gamma-3} \Gamma(\gamma) - \frac{3}{4N^2}\Big(\frac{r_0}{a}\Big)^{4\gamma-5} \Gamma(2\gamma-1)   \notag \\ &+ \frac{5}{8N^4}\Big(\frac{r_0}{a}\Big)^{6\gamma-7} \Gamma(3\gamma-2)\bigg].
\end{eqnarray}

We now compute the pair annihilation rate ($\lambda_a$) by finding the average pair lifetime. We consider a pair with maximum angle of separation $\psi = 2\theta_{max}$, and treat the insects as random walkers, a reasonable approximation since the pair forms a center manifold. Each random walker rotates by an angle $\eta \pi U[-1, 1]$ every time step, where $U$ is the uniform distribution. The variance of this distribution is $(\eta \pi)^2/3$. The pair loses stability when the angular separation exceeds $\pi/2$. Our effective diffusion is half the variance of the steps. However, since their are two particles, the diffusion constant of their difference is twice that of a single particle. So we have diffusion constant, $D_{diff} = (\eta \pi)^2/3$.

Using the first-passage time relationship, we compute the expected time it takes the pair to reach a separation of $\pi/2$ from starting angle, $\psi$:

\begin{eqnarray} 
\mathbf{E}(\tau, \psi) = \frac{(\pi/2)^2 - \psi^2}{2D_{diff}} = \frac{(\pi/2)^2 - \psi^2}{2(\eta \pi)^2/3}.
\end{eqnarray}

\noindent We then integrate over all possible starting angles, assuming that all are equally likely:

\begin{eqnarray} 
\mathbf{E}(\tau) = \frac{3}{2(\eta \pi)^2} \int_0^{\pi/2} \Big(\frac{\pi^2}{4} - \psi^2 \Big) d\psi   = \frac{\pi}{8 \eta^2}.
\end{eqnarray}

\noindent This tells us the expected number of time steps for a pair to reach a maximum separation of $\pi/2$ and break up. Expressing this as a rate, we find

\begin{eqnarray} 
\lambda_a = \frac{1}{\mathbf{E}(\tau) \Delta t} = \frac{v_0}{\mathbf{E} (\tau)\ell } = \frac{8 v_0 \eta^2}{\pi \ell}.
\end{eqnarray}

\noindent Finally, we equate the formation and annihilation rates ($\lambda_f = \lambda_a$) to find the bifurcation curve separating the stable swarm regime from the stable pair regime:

\begin{eqnarray} 
\eta_c &= \sqrt{\frac{\ell}{32\pi r_0}}  \Bigg[\Big(\frac{r_0}{a}\Big)^{2\gamma-3} \Gamma(\gamma) - \frac{3}{4N^2}\Big(\frac{r_0}{a}\Big)^{4\gamma-5} \Gamma(2\gamma-1)   \notag \\ &+ \frac{5}{8N^4}\Big(\frac{r_0}{a}\Big)^{6\gamma-7} \Gamma(3\gamma-2)\Bigg]^{\frac{1}{2}}.
\end{eqnarray}

\noindent We plot this bifurcation curve and find good agreement with numerical simulations. As defined earlier, $a = R_s\sqrt{2\log(2)} \approx 1.18 R_s$. We approximate the swarm radius as $r_0 = 2R_s \approx 11.28$.

\clearpage

\section*{Supplementary material}

\section{Justification for model assumptions}

Here we provide justification for each of the assumptions made in developing the numerical model. The assumptions are based on data from six swarms of male \textit{Anopheles coluzzii} taken from \cite{podaSpatialTemporalCharacteristics2024}, as discussed in the main text. Throughout this section, the six distinct colors correspond to the six swarming events during peak activity.

\begin{figure}[h!]
\includegraphics[width=\columnwidth]{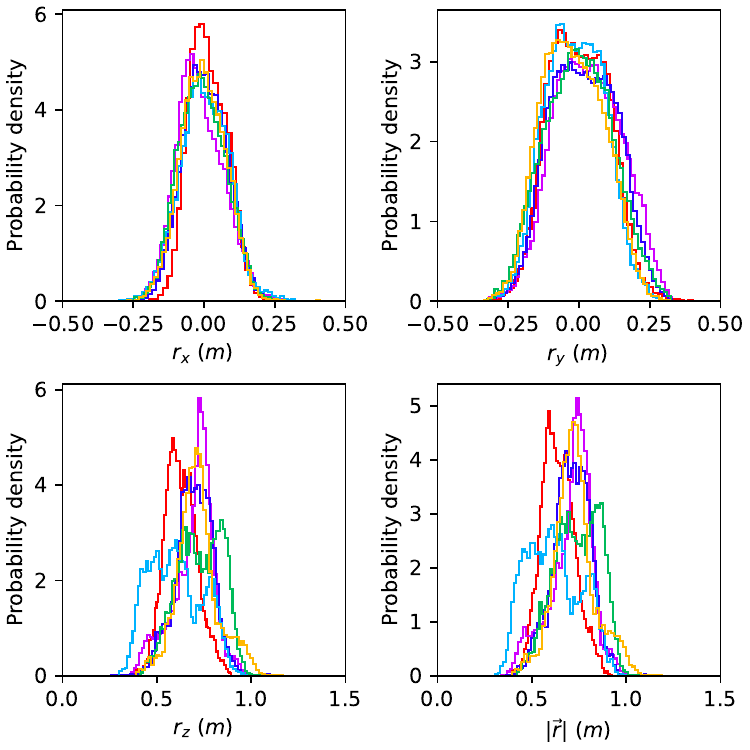}
\caption{Positional histograms over all time and all insects for each of the six swarms. We show individual components of the position vector, $\vec{r}$, as well as its magnitude. The origin is defined as the center of the swarm marker.}
\label{fig-position}
\end{figure}

\begin{figure}[h!]
\includegraphics[width=\columnwidth]{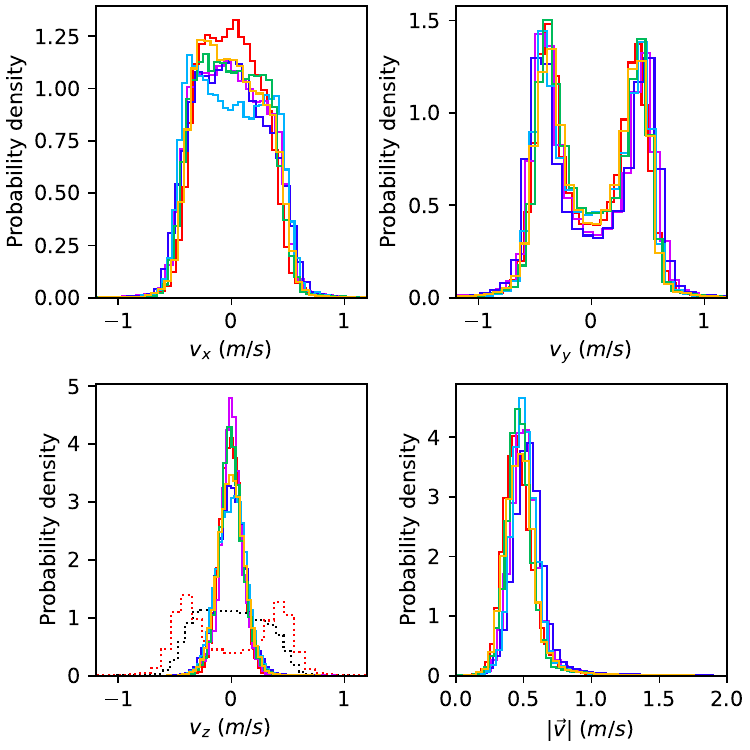}
\caption{Velocity histograms over all time and all insects for each of the six swarms. We show individual components of the velocity vector, $\vec{v}$, as well as its magnitude. For comparison, in the $v_z$ panel, we plot the average $v_x$ and $v_y$ histograms with black and red dotted curves, respectively.}
\label{fig-velocity}
\end{figure}

\begin{figure}[h!]
\includegraphics[width=\columnwidth]{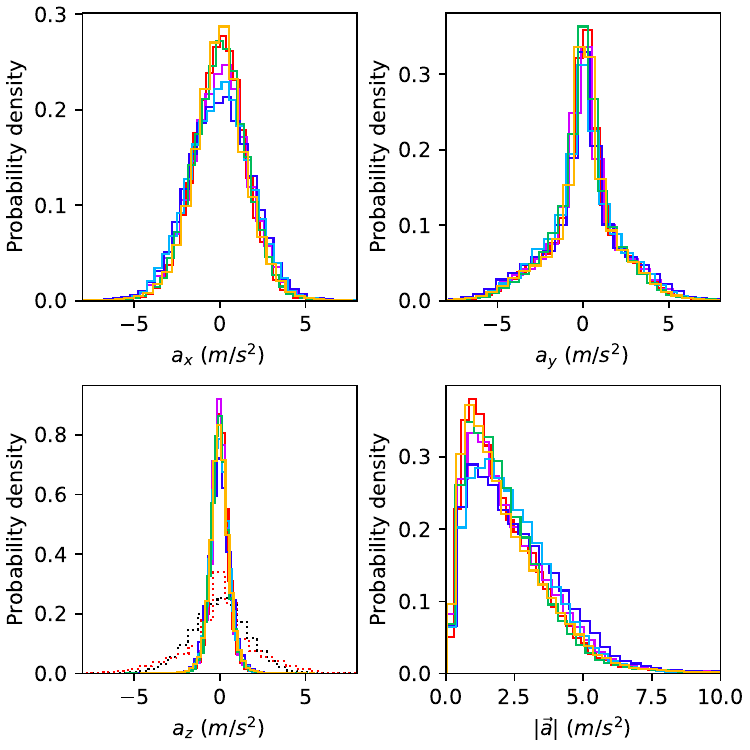}
\caption{Acceleration histograms over all time and all insects for each of the six swarms. We show individual components of the acceleration vector, $\vec{a}$, as well as its magnitude. For comparison, in the $a_z$ panel, we plot the average $a_x$ and $a_y$ histograms with black and red dotted curves, respectively.}
\label{fig-acceleration}
\end{figure}

In FIGS. \ref{fig-position}-\ref{fig-acceleration}, we show the probability distributions of insect position, velocity, and acceleration, respectively. We see that the insects fill all three spacial dimensions and exhibit what resemble Gaussian distributions in position. However, the velocity and acceleration distributions show a strong preference for dynamics in the $xy$-plane, that is the plane orthogonal to the direction of gravity. For this reason, we constrain our numerical model to two spacial degrees of freedom. Further, we see that the variation in speed is small, justifying the constant-speed assumption in our numerical model.

\clearpage

\begin{figure}[h!]
\includegraphics[width=\columnwidth]{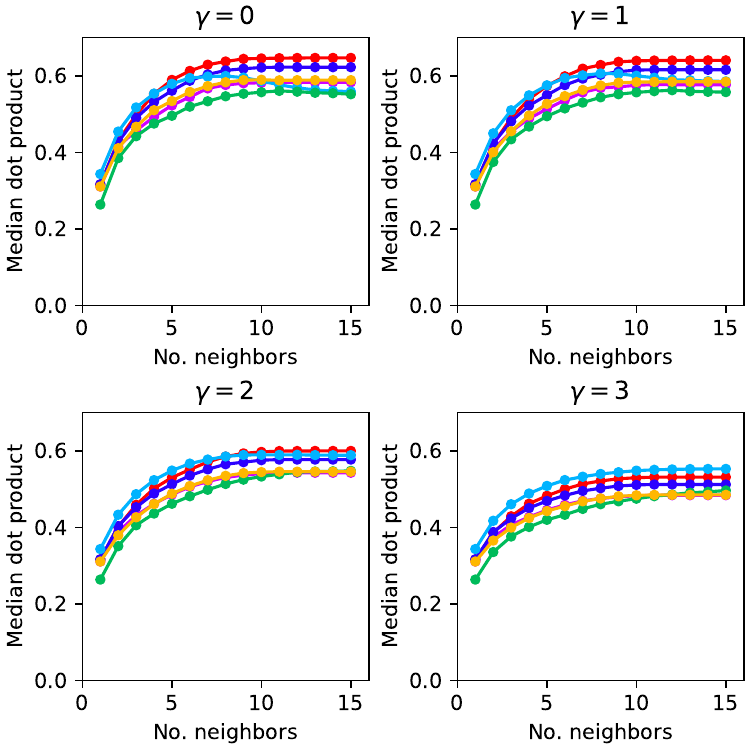}
\caption{Dot product between the predicted direction of acceleration, $\hat{D}_i$ and the measured direction of acceleration, $\hat{a}$. We vary the number of neighbors considered in the weighted mean field. Each point represents the median value across all insects and all time for a particular swarm.}
\label{fig-neighbor_count}
\end{figure}

Next, we plot statistical measures of the acceleration and position data that justify our choice of interaction rules in the numerical model. First, as a rough estimate of predictive power, we compute the dot product between the direction of the weighted mean field, $\hat{D}_i$, and the experimentally measured direction of acceleration for every point in time and all insects (FIG. \ref{fig-neighbor_count}). For perfect predictive power, the dot product takes a value of $1$. We see that this dot product increases when more neighbors are considered in the weighted mean field, regardless of $\gamma$. The dot product plateaus as the number of neighbors considered approaches the size of the full swarm. This observation justifies our choice of having an attractive force to the weighted mean field of all neighbors, rather than just a few. 

\begin{figure}[h!]
\includegraphics[width=\columnwidth]{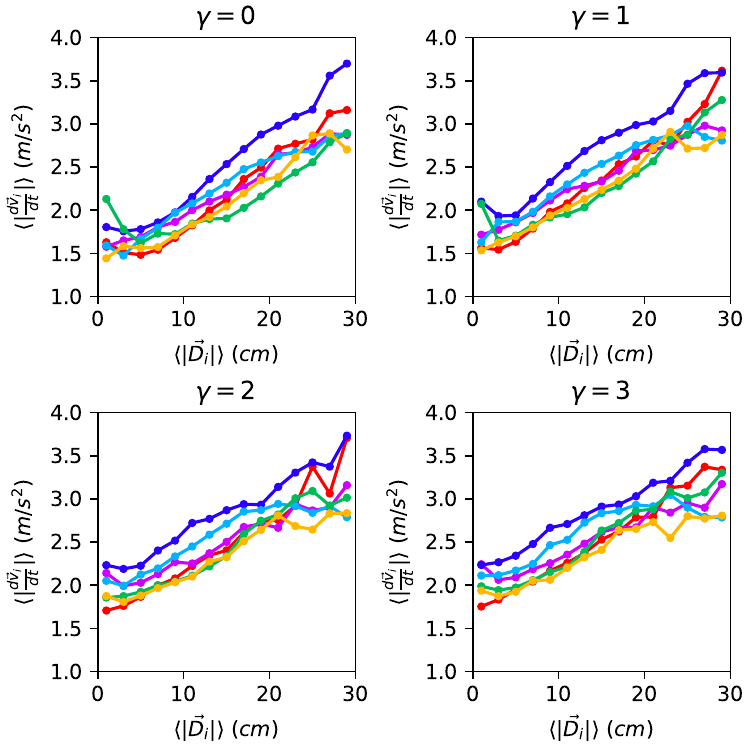}
\caption{Magnitude of measured acceleration vectors as a function of an insect's distance from its weighted mean field. For all swarms and all choices of $\gamma$, we see a nearly linear dependence.}
\label{fig-attraction}
\end{figure}

We also plot the magnitude of the measured acceleration vector as a function of an insect's distance from the weighted mean field (FIG. \ref{fig-attraction}). Regardless of our choice for $\gamma$, we find an approximately linear dependence of the acceleration magnitude on the distance from the weighted mean field. This justifies the linear attractive force incorporated in our model.

\clearpage

\begin{figure}[h!]
\includegraphics[width=\columnwidth]{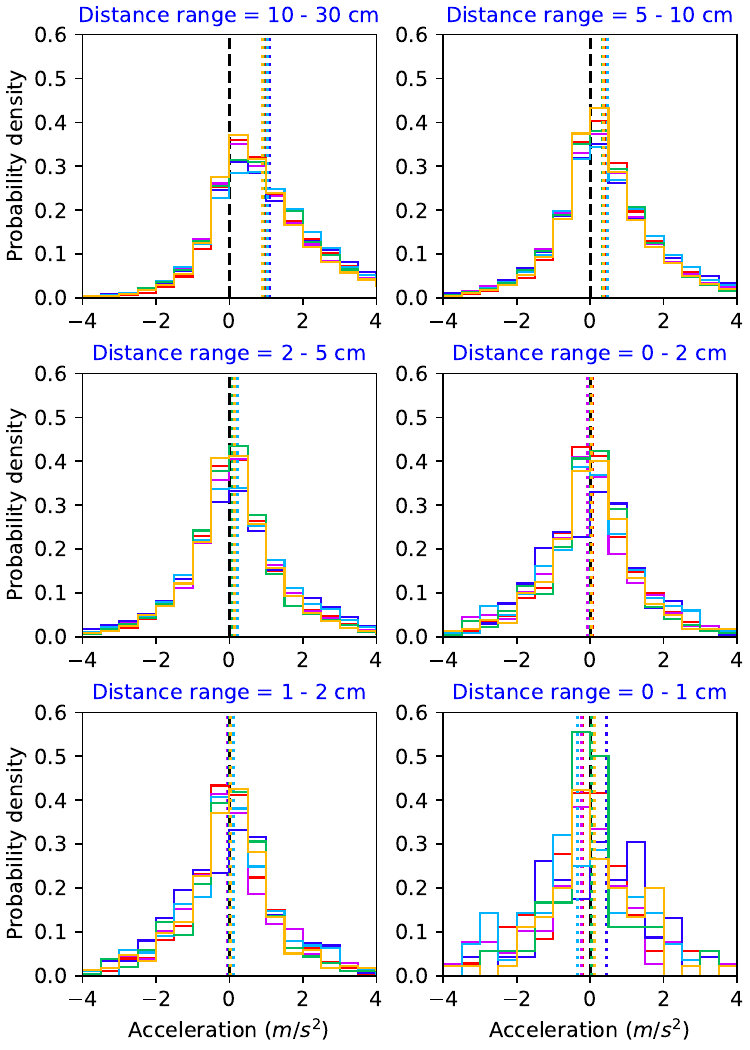}
\caption{Histograms of experimentally measured acceleration in the direction of each mosquito's nearest neighbor. Data are grouped into various ranges of distance from the nearest neighbor. Vertical, dotted lines indicate the means of the distributions. Data become sparse in the 0--1 cm regime, with each distribution containing fewer than 100 samples.}
\label{fig-repulsion}
\end{figure}

Next, we show the statistics of the experimentally measured acceleration vectors projected onto the direction of each mosquito's nearest neighbor. If a short-range repulsive force is present, we should expect these scalar projections to be biased in the negative direction. However, at short ranges, we find distributions that are nearly symmetric around zero (FIG. \ref{fig-repulsion}).

\newpage

\begin{figure}[h!]
\includegraphics[width=\columnwidth]{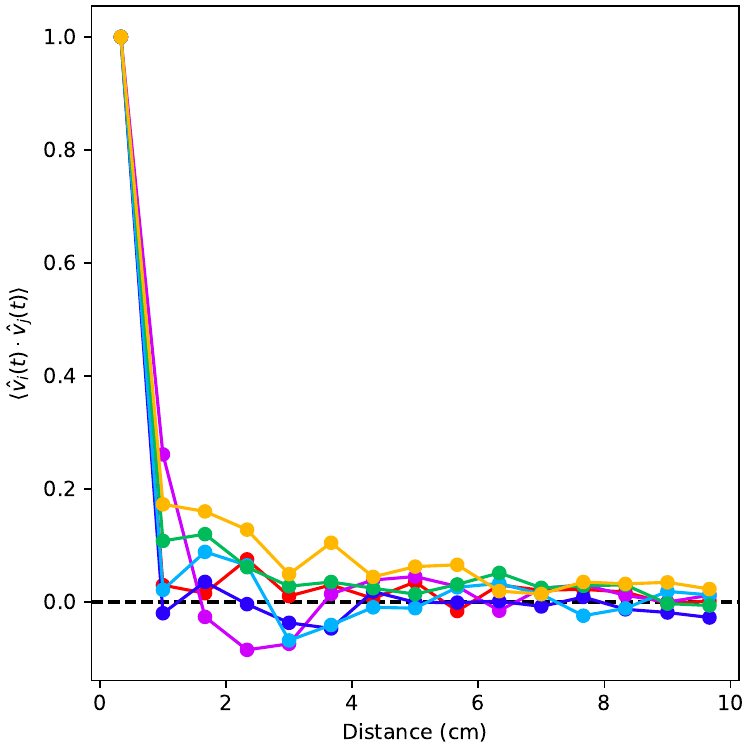}
\caption{Time-average velocity alignment between all pairs of mosquitoes. A weak bias towards positive alignment can be seen at short distances. We include each individual's alignment with itself, which is $1$ by definition.}
\label{fig-alignment1}
\end{figure}

We now search for velocity alignment correlations between pairs of mosquitoes in the experimental data. We compute the dot product between velocity unit vectors as a function of distance between the two mosquitoes. We first plot the time-average velocity alignment over all insect pairs for the six swarms (FIG. \ref{fig-alignment1}). A weak bias towards positive alignment can be seen at short distances.

\newpage

\begin{figure}[h!]
\includegraphics[width=\columnwidth]{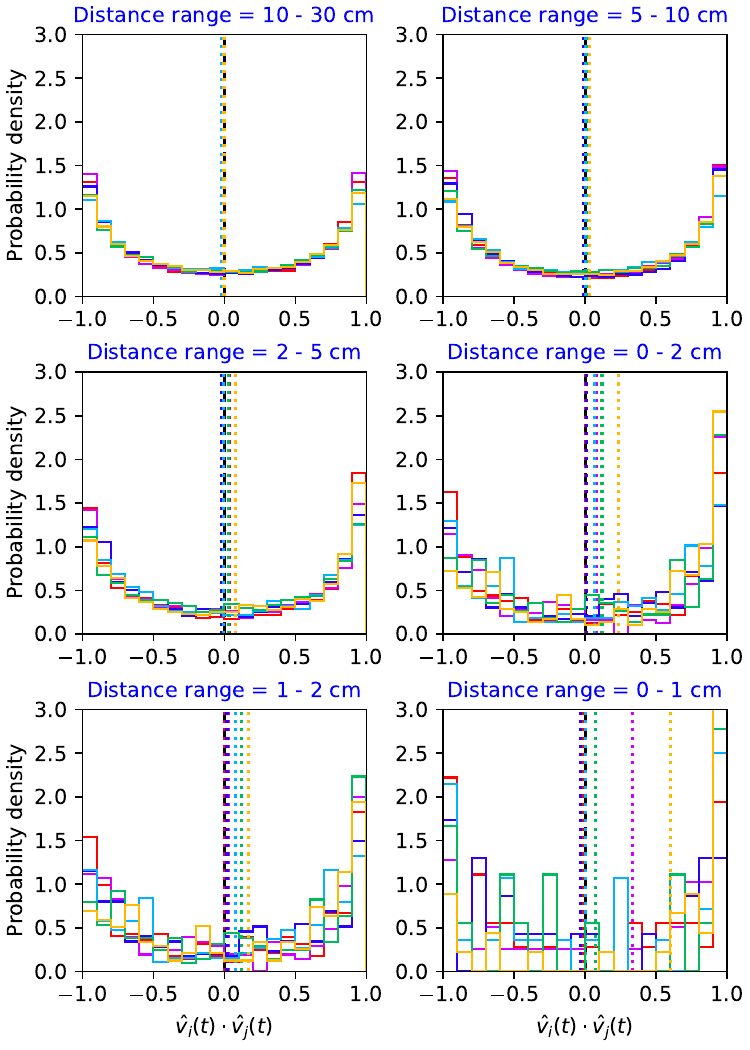}
\caption{Histograms of velocity alignment between each mosquito and its nearest neighbor. Data are grouped into several ranges of distance from the nearest neighbor. Vertical, dotted lines indicate the means of the distributions. Data become sparse in the 0--1 cm regime, with each distribution containing fewer than 100 samples.}
\label{fig-alignment2}
\end{figure}

We also show the full histograms of the velocity dot product between each insect and its nearest neighbor only (FIG. \ref{fig-alignment2}). We would expect uniform histograms of the velocity dot product if flight paths were uncorrelated random walks. However, we find that values near $\pm 1$ are much more probable  that values near zero. We attribute this bias to the circle-like trajectories the mosquitoes follow in both the clockwise and anticlockwise directions, as viewed from above.

\newpage

\section{Mapping data onto the model}

We now attempt to map the experimental data onto the numerical model. We first set $N=20$, approximating the number of mosquitoes in the experimental swarms. We then seek to find the optimal values of $\gamma$ and $\eta$, such that the model swarm most closely resembles the data. We use two dimensionless measures of insect trajectories to compare the datasets to our dimensionless model.

\begin{figure*}[h!]
\includegraphics[width=\textwidth]{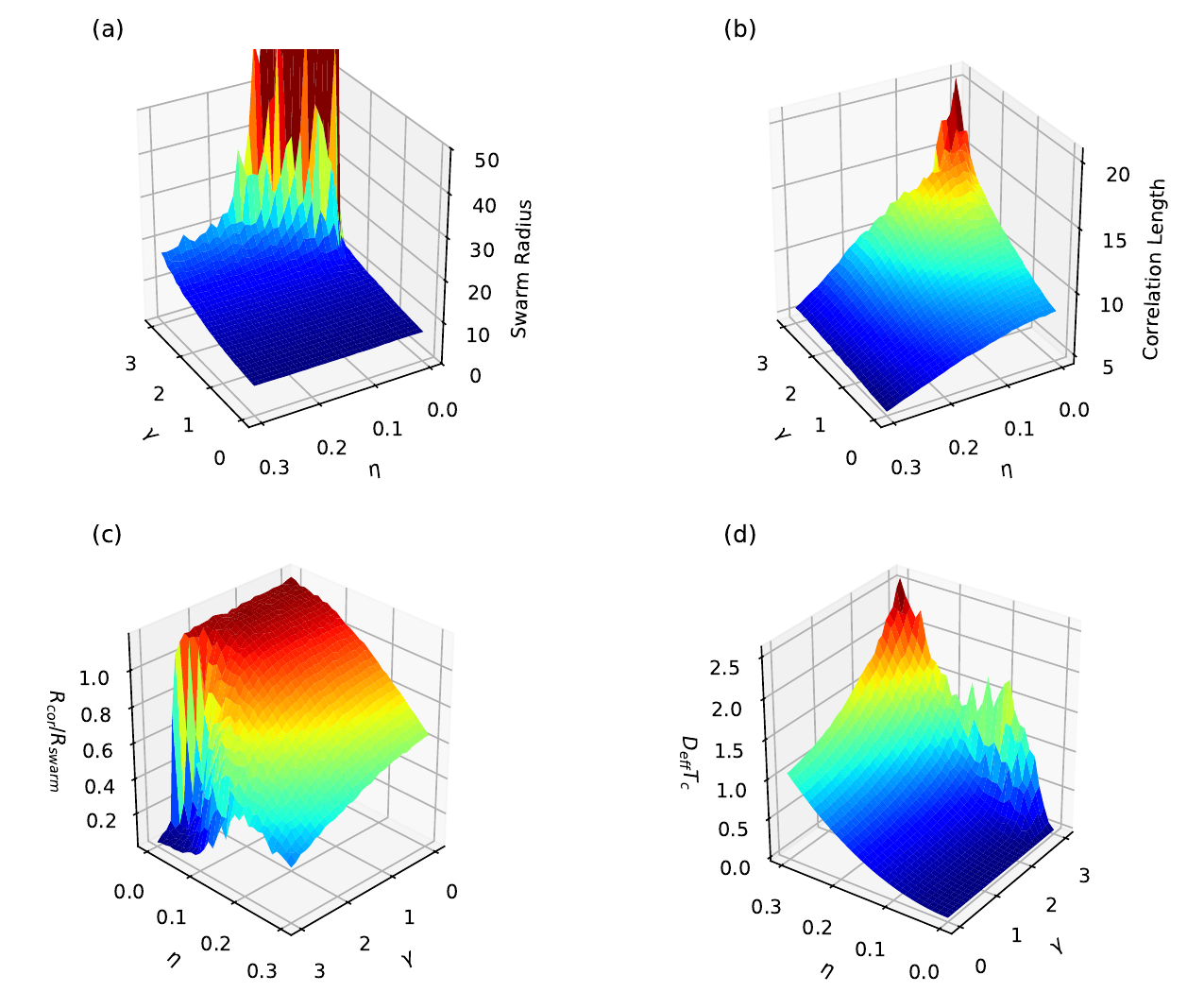}
\caption{(a) mean swarm radius of the numerical model as a function of $\gamma$ and $\eta$. The radius diverges for low values of $\eta$ and high values of $\gamma$, as the swarm becomes unstable. (b) mean correlation length of individual insect trajectories. (c) Dimensionless correlation length as defined by the ratio of the correlation length to the mean swarm radius. (d) dimensionless angular diffusion constant as a function of $\gamma$ and $\eta$.}
\label{fig-mapping1}
\end{figure*}

Our first measure is the scaled correlation length of individual trajectories. We define the correlation time, $\tau_c$, as the average time it takes for the individual trajectories to de-correlate,

\begin{eqnarray}
\mathbf{E} \big[ \hat{v}_i(t) \cdot \hat{v}_i(t + \tau_c) \big] = e^{-1},
\end{eqnarray}

\noindent where $\mathbf{E}$ is the expected-value operator. To get the dimensionless correlation length, we multiply $\tau_c$ by the mean insect speed and divide by the mean swarm radius. We show these quantities for the numerical model in FIG. \ref{fig-mapping1}(a-c). For confined Brownian motion of swarm members, we should expect a dimensionless correlation length of zero. Highly correlated motion of swarm members should yield values on the order of $1$.

Our second dimensionless measure of swarming is the dimensionless angular diffusion constant, which captures correlation in individual trajectories on short time scales. We use the angular difference, $\Delta \theta(t)$, between an insect's velocity vector at time $t$ and time $0$, and make the approximation,

\begin{eqnarray}
\langle \Delta \theta (t)^2 \rangle \approx  2D_{eff} t,
\end{eqnarray}

\noindent where $D_{eff}$ is the effective angular diffusion constant. For the experimental data, we compute the angular change over just one time step and compute $D_{eff}$. We define the dimensionless angular diffusion constant to be $D_{eff}T_c$, where $T_c$ is the characteristic time scale, defined by the ratio of the mean swarm radius to the mean speed. For the numerical model, the dimensionless diffusion constant takes the form,

\begin{eqnarray}
D_{eff}T_c = \frac{(\pi\eta)^2}{6}\bigg(\frac{R_s}{\ell} \bigg),
\end{eqnarray}

\noindent where $R_s$ is the mean swarm radius. In FIG. \ref{fig-mapping1}(d), we show how the dimensionless angular diffusion constant varies in the parameter space of the numerical model. As expected, it increases with increasing $\gamma$ and $\eta$.

\clearpage

\begin{figure}[h!]
\includegraphics[width=\columnwidth]{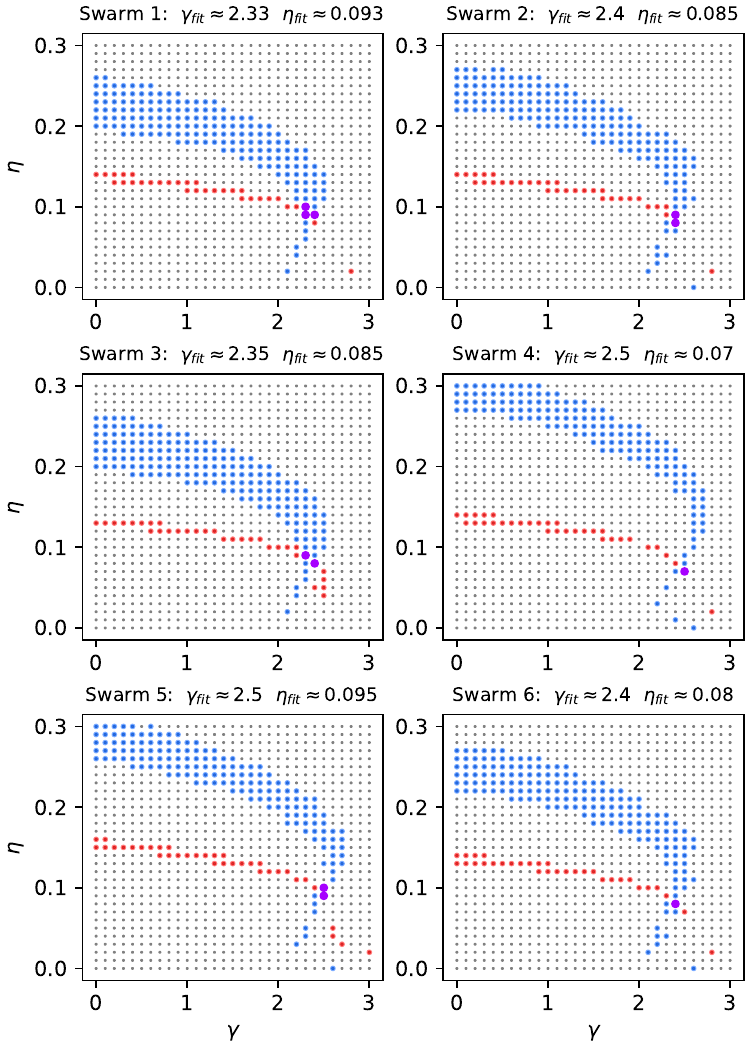}
\caption{Parameter values of agreement between data and model for all six swarms. Blue points show the parameter values that yield dimensionless correlation lengths within 10\% of the value computed for the swarm data. Red points show where the dimensionless diffusion constant is within 10\% of that of the data. Purple points show the intersection of these two regimes, where both measures are within 10\% of the experimental values. Experimental position data were filtered with a 3rd-order, 5-point Savitzky-Golay filter. If multiple intersection points were found, we simply take the mean of their values. These values are plotted in the main text.}
\label{fig-mapping2}
\end{figure}

\begin{figure}[h!]
\includegraphics[width=\columnwidth]{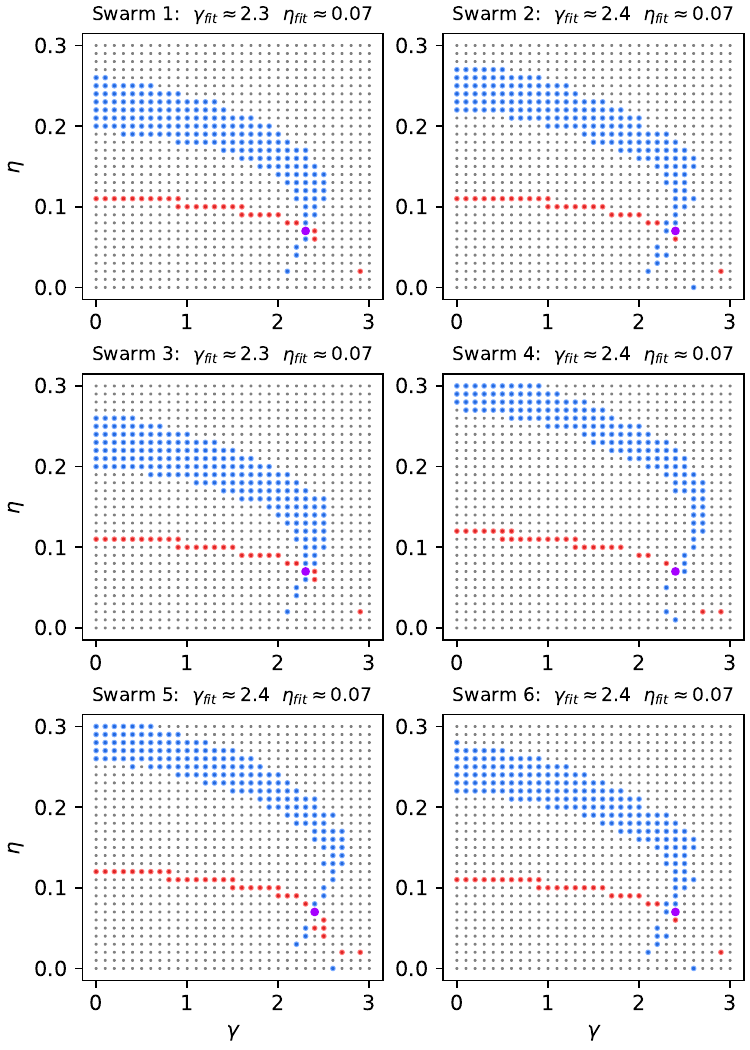}
\caption{Parameter values of agreement between data and model for all six swarms. Blue points show the parameter values that yield dimensionless correlation lengths within 10\% of the value computed for the swarm data. Red points show where the dimensionless diffusion constant is within 10\% of that of the data. Purple points show the intersection of these two regimes, where both measures are within 10\% of the experimental values. Experimental position data were filtered with a 3rd-order, 15-point Savitzky-Golay filter.}
\label{fig-mapping3}
\end{figure}

We now compute the two dimensionless measures and compare to those of the numerical model. We show the model parameter values for which the model and the data agree within 10\%. Since the diffusion constant at short time scales will likely depend on how much smoothing the data undergo, we vary the window of the Savitzky-Golay filter. We show the mapping for a narrow, 5-point window (FIG. \ref{fig-mapping2}), which provides minimal smoothing. We then show the same results using our standard 15-point filter (FIG. \ref{fig-mapping3}). Varying the filter window has only a small effect on the best-fit values and shows no systematic trend.

\clearpage

\section{Relationship between $L$, $N$, and mean swarm radius}

\begin{figure}[h!]
    \centering
    \includegraphics[width=\linewidth]{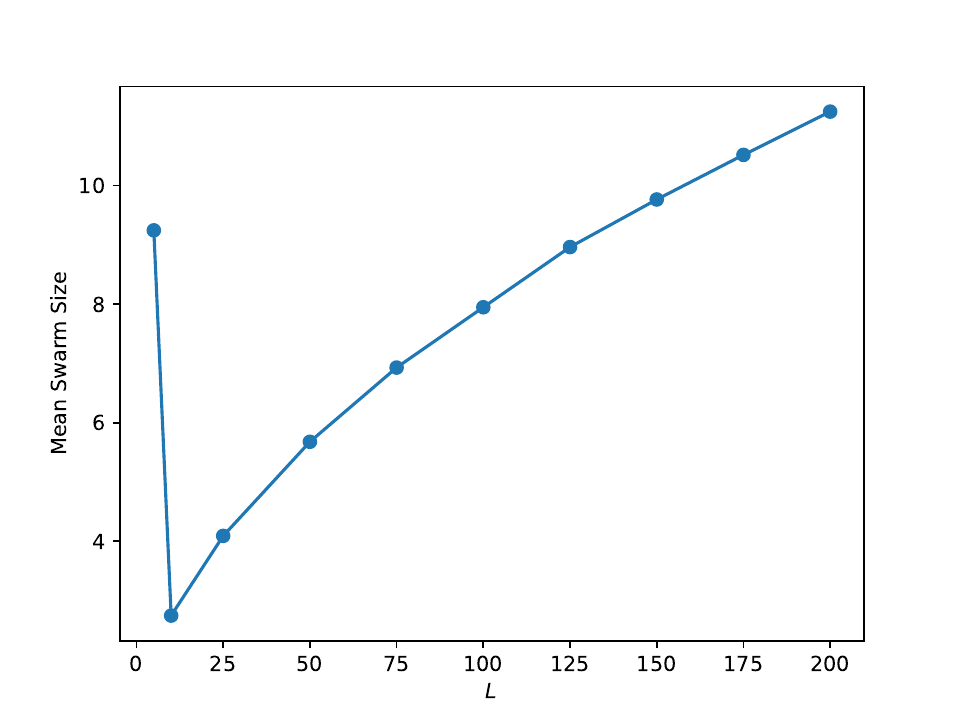}
    \caption{Mean swarm size as a function of $L$ demonstrating the $R_s \propto \sqrt{L}$ relationship found in the main text. We set $N=100$, $\gamma = 0.0$, and $\eta = 0.5$.}
    \label{fig:mss_L}
\end{figure}

\begin{figure}[h!]
    \centering
    \includegraphics[width=1\linewidth]{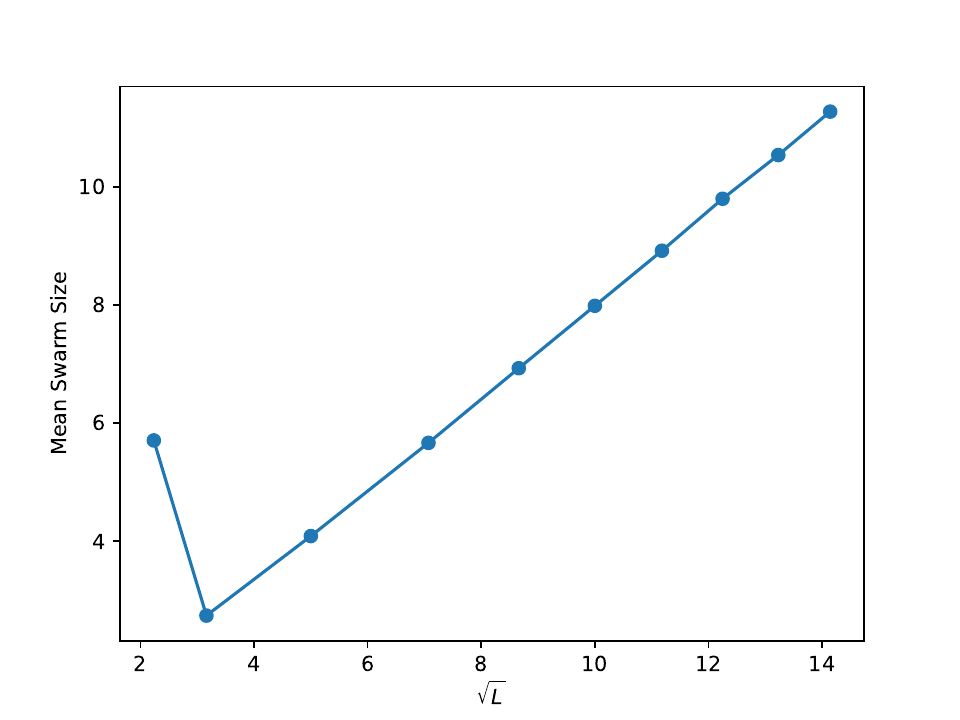}
    \caption{Mean swarm size as a function of $\sqrt{L}$, confirming the $R_s \propto \sqrt{L}$ relationship shown in the previous figure.}
    \label{fig:mss_rootL}
\end{figure}

In this section, we show how the mean swarm size, $R_s$, depends on $L$ and $N$. We first look at the relationship between $R_s$ and $L$. In FIG. \ref{fig:mss_L}, we show that $R_s$ is proportional to $\sqrt{L}$, consistent with the \textit{Length scale of swarm} calculation in the main text. In FIG. \ref{fig:mss_rootL}, we plot $R_s$ as a function of $\sqrt{L}$ and confirm that the relationship is linear.

\newpage

\begin{figure}[h!]
    \centering
    \includegraphics[width=\linewidth]{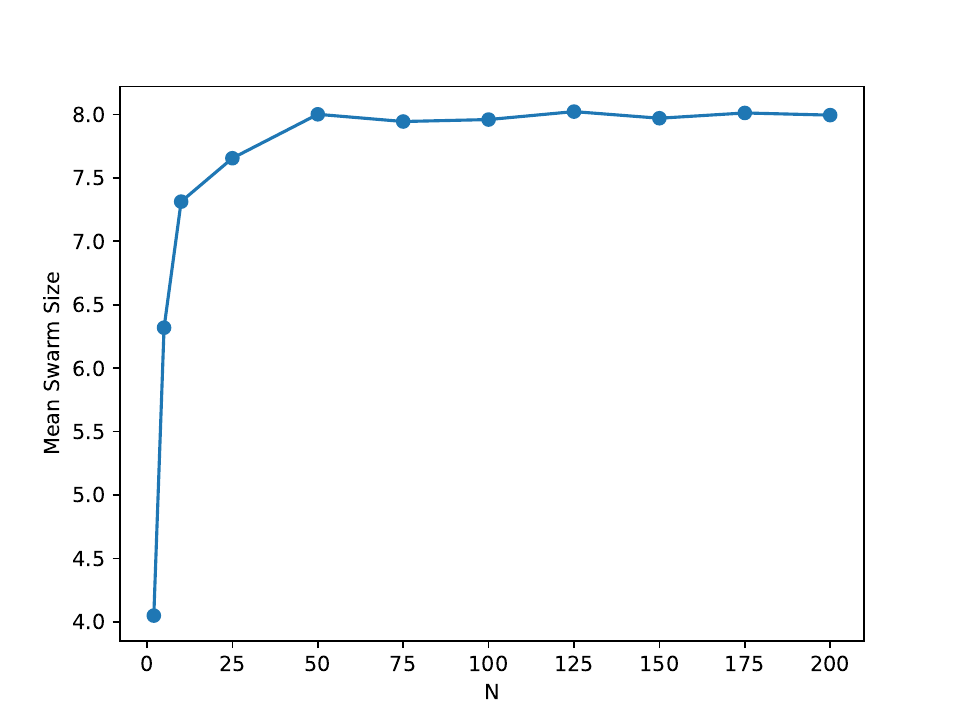}
    \caption{Mean swarm size as a function of $N$. The mean swarm size plateaus at $N \approx 50$. We set $L=100$, $\gamma = 0.0$, and $\eta = 0.5$.}
    \label{fig:mss_N}
\end{figure}

Next, we look at the relationship between $R_s$ and $N$. FIG. \ref{fig:mss_N} shows that the mean swarm size initially increases with $N$. At $N \approx 50$, the mean swarm size plateaus. 

\begin{figure}[h!]
    \centering
    \includegraphics[width=\linewidth]{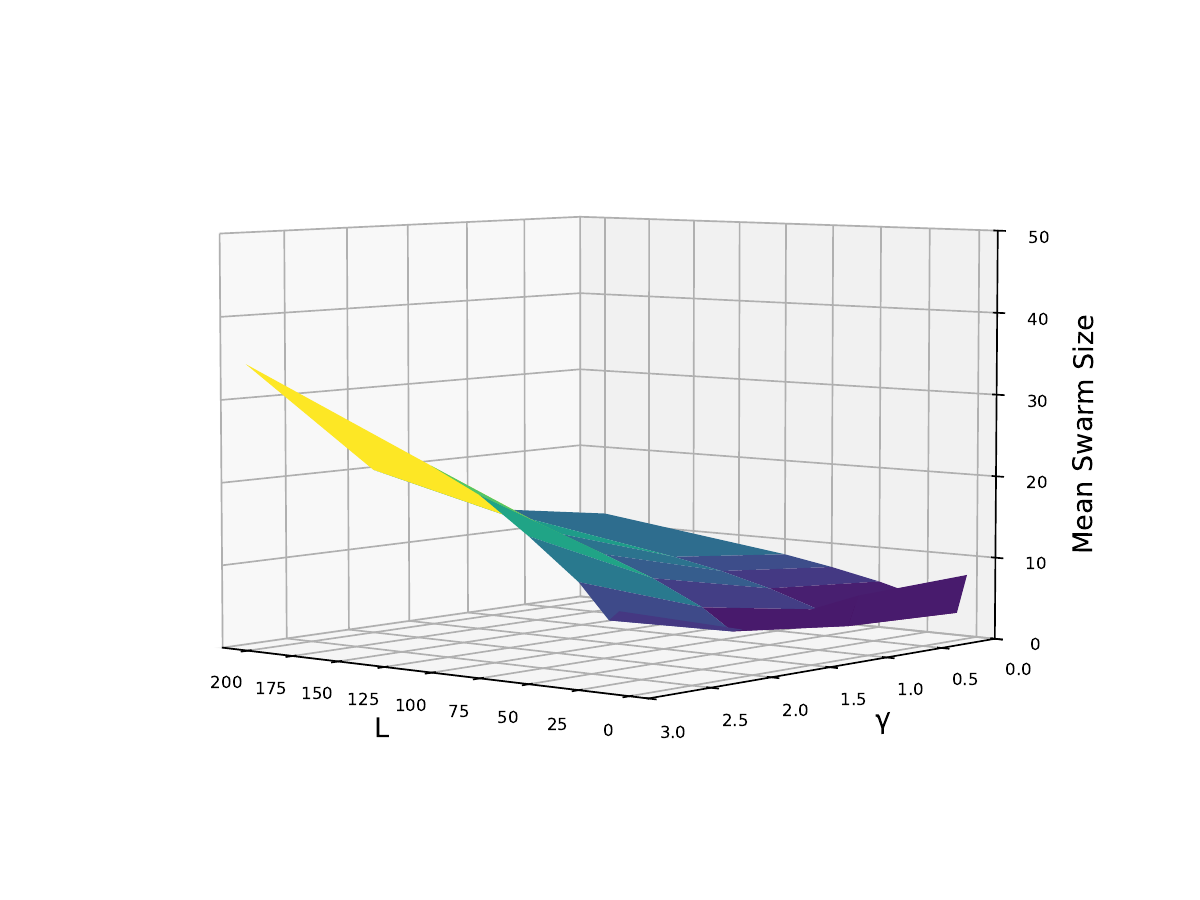}
    \caption{Surface relating mean swarm size, $L$, and $\gamma$. $\eta$ was held constant at 0.5.}
    \label{fig:gamma_surface}
\end{figure}

We now examine the relationship between the mean swarm size and $L$ over a range of values for $\gamma$ and $\eta$. These relationships are shown using surface plots. In FIG. \ref{fig:gamma_surface}, we sweep $\gamma$ from 0.0 to 3.0. We see that the mean swarm size increases with $L$ for $\gamma$ in this range as expected.

\newpage

\begin{figure}[h!]
    \centering
    \includegraphics[width=1\linewidth]{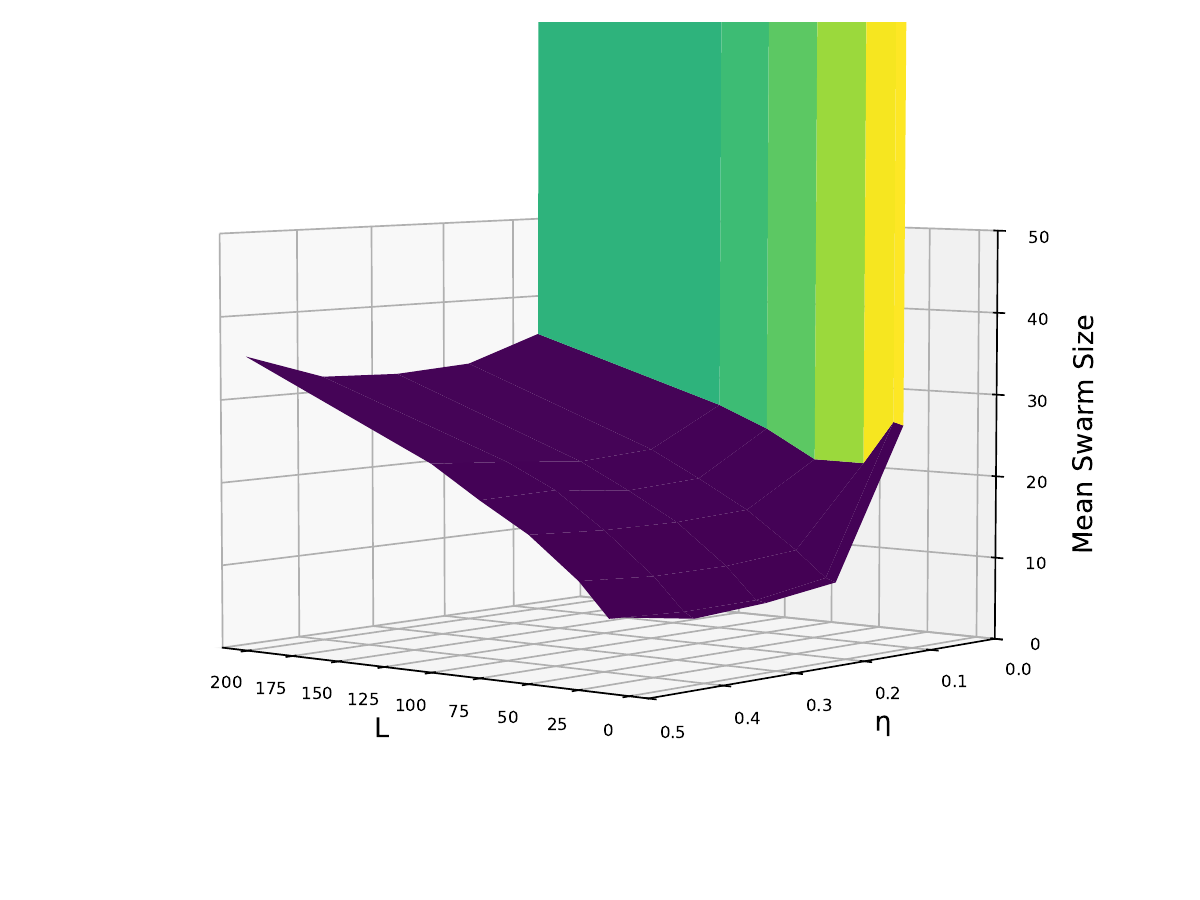}
    \caption{Surface relating mean swarm size, $L$, and $\eta$. The swarm size blows up at low $\eta$ for all values of $L$ as expected. For this surface, $\gamma$ was held constant at 3.}
    \label{fig:eta_surface}
\end{figure}

In FIG. \ref{fig:eta_surface}, we plot the mean swarm size as a function of $L$ and $\eta$, with $\eta$ values ranging from 0.0 to 0.5. The swarm becomes unstable at small values of $\eta$, when there is little to no noise.

\newpage

\section{Estimating the upper bifurcation curve}

We now approximate the upper bifurcation curve, which is plotted with green curves in the main text. This phase transition occurs in the high noise regime ($\eta > 0.5$), where pairs are quickly destabilized by noise. For high enough noise, the whole swarm slowly disintegrates. Near the swarm center, the trajectories can be approximated as random walks. However, when far from the swarm center, the mean field imposes a bias, and the insects can be regarded as biased random walkers. We want to estimate how far from the swarm center an insect needs to be for this bias to kick in. Based on the geometry of the problem, we can approximate this distance to be,

\begin{eqnarray} 
r_d = \frac{r_0}{\sin\Big((1-\eta)\pi\Big)}.
\end{eqnarray}

\noindent We will refer to this quantity as the diffusion radius. Inside the diffusion radius, the bias is removed and the insects interact in a diffusive manner, with all orientation angles almost equally likely with every time step. Outside the diffusion radius, the collective attraction to swarm members does not cancel and an attractive bias towards the swarm center can be found.

Next, we find the radius at which an insect become more attracted to a single neighbor than a swarm of $N$ other insects. We assume the single insect and the swarm of $N$ insects are on opposing sides of our reference insect. Using Eq. 4 from the main text and setting $\vec{D}=0$, we find

\begin{eqnarray} 
N\frac{r}{r^\gamma} - \frac{a}{a^\gamma} = 0,
\end{eqnarray}

\noindent where $a$ is the interaction radius defined in the main text. Solving for the critical radius at which the two forces are equal and opposite, we find

\begin{eqnarray} 
r_c = a N^{\frac{1}{\gamma-1}}.
\end{eqnarray}

\noindent We now set $r_c = r_d$ and solve for the critical noise strength in order to get the upper bifurcation curve,

\begin{eqnarray} 
\eta_{c2} = 1 - \frac{1}{\pi}\arcsin\Big(\frac{r_0}{a} N^{ \frac{1}{1 - \gamma}} \Big),
\end{eqnarray}

\noindent which is valid only for $\eta > 0.5$. As in the calculations from the main text, we use $a = R_s\sqrt{2\log(2)} \approx 1.18 R_s$ and $r_0 = 2R_s \approx 11.28$.

\clearpage

\end{document}